\documentclass[a4paper,12pt,twoside,notitlepage]{article}
\usepackage[english]{babel}
\usepackage{a4,epsfig,latexsym,amssymb}
\usepackage{amsmath,amssymb,graphicx,makeidx,mathrsfs}
\usepackage{axodraw}
\newcommand{\bei}{\begin{itemize}}
\newcommand{\eei}{\end{itemize}}
\newcommand{\beq}{\begin{equation}}
\newcommand{\eeq}{\end{equation}}
\newcommand{\beqn}{\begin{eqnarray}}
\newcommand{\eeqn}{\end{eqnarray}}
\newcommand{\beqns}{\begin{eqnarray*}}
\newcommand{\eeqns}{\end{eqnarray*}}

\newcommand\Amptpbar{\kern 0.18em\overline{\kern -0.18em {\cal A}}_{3\pi}}

\newcommand\Amptpbarkappa{\kern 0.18em\overline{\kern -0.18em A}^{\kappa}{}}
\newcommand\Amptpbarsigma{\kern 0.18em\overline{\kern -0.18em A}^{\sigma}{}}

\newcommand\Nbpm{{\kern 0.18em\overline{\kern -0.18em N}}^{+-}}
\newcommand\Nbmp{{\kern 0.18em\overline{\kern -0.18em N}}^{-+}}

\newcommand\BRpmb{{\cal \kern 0.18em\overline{\kern -0.18em  B}}{}_{\rho\pi}^{+-}}
\newcommand\BRmpb{{\cal \kern 0.18em\overline{\kern -0.18em  B}}{}_{\rho\pi}^{-+}}

\newcommand\BRipmb{{\cal \kern 0.18em\overline{\kern -0.18em  B}}{}_{\rho^+\pi^-}}
\newcommand\BRimpb{{\cal \kern 0.18em\overline{\kern -0.18em  B}}{}_{\rho^-\pi^+}}

\newcommand\Abar{\kern 0.18em\overline{\kern -0.18em A}{}}

\newcommand\rfit{{\em R}fit}

\newcommand\ckmfitter{{CKMfitter}}

\newcommand{\simgt}{\,\hbox{\lower0.6ex\hbox{$\sim$}\llap{\raise0.6ex\hbox{$>$}}}\,}
\newcommand{\simlt}{\,\hbox{\lower0.6ex\hbox{$\sim$}\llap{\raise0.6ex\hbox{$<$}}}\,}

\makeatletter
\renewcommand*\env@matrix[1][c]{\hskip -\arraycolsep
  \let\@ifnextchar\new@ifnextchar
  \array{*\c@MaxMatrixCols #1}}
\makeatother

\begin{document}
\begin{titlepage}
\begin{titlepage}

\setcounter{page}{1}

{\small
\begin{flushright} 
        \today
\end{flushright} 
}
\begin{center}
\vspace{0.3cm}
{\Large\bf The Two Higgs Doublet Model of Type II\\ facing flavour physics data}\\
\vspace{0.3cm}
\vspace{0.3cm}
\end{center}

\begin{center}
{\normalsize
O.~Deschamps$^{\,c}$,
S.~Descotes-Genon$^{\,f}$,
S.~Monteil$^{\,c}$,
V.~Niess$^{\,c}$,
S.~T'Jampens$^{\,a}$,
V.~Tisserand$^{\,a}$
} 
\vspace{0.3cm}
\noindent
for the \ckmfitter\ Group. \\
\vspace{0.3cm}
\end{center}
\centerline{\small{\bf Abstract}} 
\vspace{0.1cm}
\noindent
{\small
\par
We discuss tests of the charged Higgs sector of the Two Higgs Doublet Model (2HDM) of Type II in the light of recent flavour physics data.
Particular attention is paid to recent measurement of purely leptonic decays of heavy-light mesons, which depart more or less significantly from the Standard Model (SM)
predictions. We derive constraints on the parameters of the 2HDM type II from leptonic and semileptonic $\Delta F=1$ decays as well as loop processes ($b \to s \gamma$,
$B\bar{B}$ mixing or $Z\to\bar{b}b$) sensitive to charged Higgs contributions.
\par             
The outcome of this work is that while 2HDM Type II can fit individual observable through fine-tuning schemes, in a combined analysis it does not perform better than the SM by
favouring a decoupling solution. Assuming that 2HDM Type II is realized in Nature, constraints on its parameters ($m_{H^+}$ and $\tan \beta$) are derived. A limit on the charged
Higgs mass $m_{H^+} > 316 \; {\rm GeV}$ at $95\%$~CL is obtained irrespective of the value of $\tan \beta$. This limit is dominated by the $b\rightarrow s \gamma$ branching ratio
measurement. 
\par
All results have been obtained with the \ckmfitter\ analysis package,  featuring the frequentist statistical approach \rfit\ to handle theoretical uncertainties.
}

\begin{center}
 
{\small \em $^{a}$Laboratoire d'Annecy-Le-Vieux de Physique des Particules \\
                   9 Chemin de Bellevue, BP 110, F-74941
                   Annecy-le-Vieux Cedex, France\\
                   (UMR 5814 du CNRS-IN2P3 associ\'ee \`a
                   l'Universit\'e de Savoie) \\
                {e-mail: tisserav@lapp.in2p3.fr, tjamp@lapp.in2p3.fr}} \\[0.5cm]
{\small \em $^{c}$Laboratoire de Physique Corpusculaire de Clermont-Ferrand \\
                  Universit\'e Blaise Pascal\\
                  24 Avenue des Landais F-63177 Aubiere Cedex \\
		  (UMR 6533 du CNRS-IN2P3 associ\'ee \`a
                   l'Universit\'e Blaise Pascal) \\
		  {e-mail: odescham@in2p3.fr, monteil@in2p3.fr, niess@in2p3.fr}} \\[0.5cm]
{\small \em $^{f}$Laboratoire de Physique Th\'eorique \\
                   B\^{a}timent 210, Universit\'e  Paris-Sud 11, F-91405 Orsay Cedex, France \\
                   CNRS/Univ. Paris-Sud 11 (UMR 8627)\\
                {e-mail: Sebastien.Descotes-Genon@th.u-psud.fr}}\\[0.5cm]

		
\end{center}

\vspace{\stretch{1}}

\hrule\vspace{0.1cm}
{\small\noindent http://ckmfitter.in2p3.fr
      \hfill}

\newpage

\thispagestyle{empty}

\end{titlepage}

\end{titlepage}
\vspace{1.cm}

\newpage

$ ^{ } $

\newpage


\section{Introduction}

\begin{figure}[h]
\begin{center}
\epsfig{file=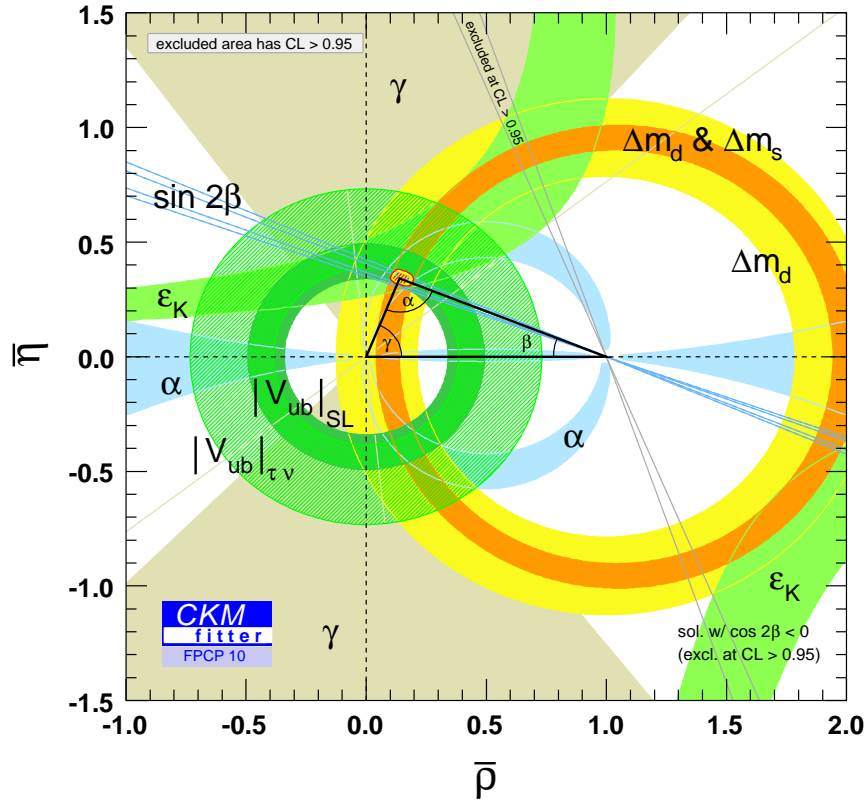,height=120mm,width=120mm}
\caption{\it \small Superimposed individual constraints at 95\% CL for the SM global fit. The yellow bean is the solution driven by the combination of all individual constraints
at 95\% CL. Regions outside the red circle are excluded at 95\% CL.\label{ckm09}} 
\end{center}
\end{figure}

The overall agreement between the Standard Model (SM) and data in the quark sector is particularly impressive: Flavour-Changing Neutral Currents (FCNC) are as small as predicted
in the SM and the Kobayashi-Maskawa (KM) mechanism has been proven to describe the observed CP-violating phenomena in flavour physics with a good accuracy. One example of this
convergence is provided by the global fit of the CKM matrix elements~\cite{SM_CKM} illustrated in Figure~\ref{ckm09}. All constraints (either loop and tree observables or
CP-violating and CP-conserving quantities) point towards a unique solution, which proves the CKM mechanism to be at work in flavour transitions (within the present accuracy) and
establishes the KM mechanism as a dominant source of CP violation in $K$- and $B$-meson systems. The consistency between these predictions is deeply related to the fact that a
single Higgs doublet provides a mass to all fermions through Yukawa couplings in the SM and that all but one CP-violating phases can be rotated away by redefinition of the fields.

Extensions of the SM are often based on the introduction of additional fields interacting with the quarks (gauge bosons of new interactions, supersymmetric particles,
technifermions~\cite{revNP}\ldots). These new fields bring along new arbitrary parameters, often inducing dangerous FCNC processes, as well as new CP-violating phases that cannot
generally be rotated away. Therefore, flavour physics and CP-violation data provide stringent constraints on the parameters of these extensions, which yield generally fine-tuning
problems. From this point a view, a rather minimal extension of the SM, with a limited number of new parameters to fix, consists in the  two-Higgs doublets models (2HDM). Indeed, in
the SM, the same doublet is used to couple the left-handed fermion doublets with right-handed up- and down-type quarks at the same time (exploiting the fact that the
representation of the Higgs doublet under $SU(2)$ is pseudoreal, and that the hypercharges of the scalar doublets coupled to right-handed up- and down-type quarks are opposite).
This choice is  imposed by no other arguments but economy, and one can introduce two complex doublets $\phi_1$ and $\phi_2$ of opposite hypercharge, rather than a single doublet
$\phi$ ~\cite{2hdmref1,2hdmref2,2hdmref3,2hdmref4,2hdmref5,2hdmref6}. One assumes that electroweak symmetry breaking occurs because the neutral components of
the two doublets acquire two {\it a priori} different vacuum expectation values: 
\begin{equation}
|\langle 0|\phi_1|0\rangle |=
   \left(\begin{array}{c} 0\\ v_1/\sqrt{2}\end{array}\right), \qquad 
|\langle 0|\phi_2|0\rangle |=
   \left(\begin{array}{c} v_2/\sqrt{2}\\ 0\end{array}\right), 
\end{equation}
denoted respectively $v_1$ and $v_2$ (with the constraint $v_1^2+v_2^2=v^2$, where $v$ is the SM Higgs vacuum expectation value).

2HDM models contain 8 degrees of freedom in the Higgs sector, out of which 3 are used to provide a longitudinal polarisation to the weak gauge bosons $W$ and $Z$. Five real fields
remain: two charged Higgs fields $H^\pm$, a neutral pseudoscalar Higgs $A$ and two neutral scalar Higgs ($h^0$ and $H^0$). The additional parameters needed to describe this SM
extension are the masses of $H^\pm$, $H^0$ and $A$, the ratio of vacuum expectation values $\tan\beta=v_2/v_1$ and an angle  describing the mixing between $h^0$ and $H^0$. 

Different versions of the 2HDM were labeled according to the couplings of the Higgs doublets to the quarks~\cite{2hdmref4}: type I corresponds to $\phi_1$ coupling to both up-
and down-type quarks whereas $\phi_2$ does not couple to any quark, type II corresponds to $\phi_1$ coupling to down-type quarks whereas $\phi_2$ couples to up-type quarks (and to
leptons), and type III to $\phi_1$ and $\phi_2$ both coupling to both types of quarks. 
Among these various possibilities 2HDM Type II is particularly alluring, because of its resemblance with the SM in the quark sector. One has two Yukawa matrices $y^{d,u}$
describing the couplings among quarks (and one for the lepton sector, $y^e$): 
\begin{equation}
\mathcal{L}_{II,Y}=-\bar{Q}_L \phi_1 y^d D_R - \bar{Q}_L \phi_2 y^u U_R - \bar{L}_L \phi_2 y^e E_R 
+h.c.,
\end{equation}
where $Q_L$ and $L_L$ denote left-handed fermion doublets, and $D_R$, $U_R$, $E_R$ down-, up-type and charged-lepton right-handed singlets, 
defined in a similar way as in the SM (actually, the SM is recovered through the identification $\phi_2=i\sigma_2\phi_1^*$, with $\sigma_2$ the complex Pauli matrix).

One has to re-express these couplings in terms of mass eigenstates. The structure of the Yukawa terms
yields a SM-like structure for the quark sector: there is a CKM matrix which is the only source of flavour-changing interactions and there are no flavour-changing neutral currents
at tree level. But there are new flavour-changing charged interactions, corresponding to the exchange of a charged Higgs rather than a $W$ (obviously, there are also interactions
of quarks with neutral Higgs fields, as well as couplings of the Higgs fields to the leptons). Once quark and Higgs fields are expressed in terms of mass eigenstates, one obtains
the following charged-Higgs interactions for quarks and leptons: 
\begin{equation}\label{HiggsLag}
\mathcal{L}_{II,H^+}=
 -\frac{g}{\sqrt{2}}\sum_{ij}
  \left[\tan\beta\frac{m_{dj}}{M_W}\bar{u}_{Li}V_{ij}d_{Rj}
        +\cot\beta\frac{m_{ui}}{M_W}\bar{u}_{Rj}V_{ij}d_{Lj}
        +\tan\beta\frac{m_{\ell i}}{M_W}\bar{\nu}_{Lj}\ell_{Rj}
           \right]H^++h.c.
\end{equation}

Therefore, the 2HDM Type II provides a very interesting extension of the SM: it exhibits the same CKM structure, has a natural mechanism of suppression of FCNC plaguing many other
models, but it exhibits a different structure for charged currents by the addition of new (scalar and pseudoscalar) interactions. 
Furthermore, it relies on a limited number of additional parameters, i.e., the mass $m_{H^+}$ of the charged Higgs and the ratio $\tan \beta$  of the couplings to up-like over
down-like quarks (if we restrict our study to charged currents). In many situations, the change induced by the 2HDM Type II amounts to a redefinition of some of the parameters
already occurring in the SM expressions, with a new dependence on $m_{H^+}$ and $\tan\beta$. 

Eventually, in addition to that predictability virtue, the 2HDM Type II is embedded into the most simple supersymmetric extensions of the SM (MSSM), at least at tree
level~\cite{2hdmref2} (for large $\tan\beta$, loop effects might lead supersymmetric theories to coincide with a 2HDM Type III rather than the 2HDM Type II). Searches for such
charged Higgs are obviously among the prospects of the LHC experiments ATLAS and CMS~\cite{2hdmsearch}. 

Since the decays mediated by a weak charged current are an excellent laboratory to search for charged Higgs boson contributing in addition to $W^{\pm}$ bosons, we have collected the measured decays potentially sensitive to contributions from charged Higgs, for which a good control of the theoretical hadronic uncertainties can be achieved. A
combined analysis of their branching ratios in the light of the 2HDM Type II is then performed within the frequentist statistical scheme developed by the CKMfitter
group~\cite{ThePapII}.  

It is convenient to categorize these observables as follows: 

\begin{enumerate}

\item The leptonic decays of mesons mediated by quark-annihilation at tree level $\Gamma[K\rightarrow\mu\nu]/\Gamma[\pi\rightarrow\mu\nu]$, ${\cal B}[D\rightarrow\mu\nu]$, ${\cal
B}[D_s\rightarrow\mu\nu]$, ${\cal B}[D_s\rightarrow\tau\nu]$ and ${\cal B}[B\rightarrow\tau\nu]$, where ${\cal B}$ stands for branching ratio and $\Gamma$ for the decay width. In
addition we also consider the strange hadronic decay of the $\tau$ lepton, $\tau \rightarrow K \nu$, through the ratio $\Gamma[\tau\rightarrow
K\nu]/\Gamma[\tau\rightarrow\pi\nu]$, which can be seen as reversed leptonic decays.  

\item The semileptonic decays $B \rightarrow D \tau \nu$, through the ratio ${\cal B}[B\rightarrow D\tau\nu]/{\cal B}[B\rightarrow D \ell \nu]$, and $K \rightarrow \pi \ell \nu$,
through the ratio ${\cal B}[ K \rightarrow \pi \mu \nu]/{\cal B}[ K \rightarrow \pi e \nu]$.  

\item The $B^0_d$ and $B^0_s$ oscillation frequencies, $\Delta m_{d}$ and $\Delta m_{s}$.

\item The $Z$ partial width into $b$ quarks, 
 $R_b =\Gamma[Z\rightarrow b\bar{b}]/\Gamma[Z\rightarrow {\rm hadrons}]$, which exhibits electroweak charged currents through $Z b \bar b$ vertex radiative corrections. 

\item The measurement of the FCNC radiative decay $b \to s \gamma$ through the ratio  ${\cal B}[\bar{B}\rightarrow X_s\gamma]/{\cal B}[\bar{B}\rightarrow X_ce\bar{\nu}]$. 

\end{enumerate}

\begin{figure}
\begin{center}
\epsfig{file=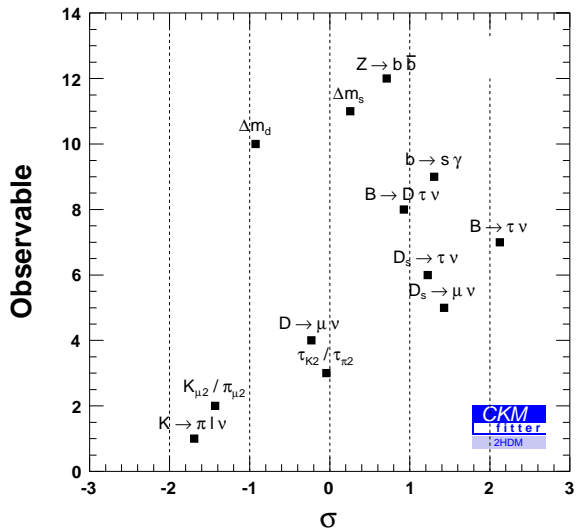,height=80mm,width=80mm}
\caption{\it \small Comparison of the measurements relevant to constrain the 2HDM Type II and their predicted value in the SM. The black dots indicate the deviation of the experimental
result from its prediction, assuming Gaussian distributed errors for both theory and experiment. The deviation is expressed as a signed significance whith positive values
indicating a mesurement higher than its prediction. Note that this figure is for illustration purposes only, since all the errors are treated as Gaussian here. In the rest of the paper, 
we use the \rfit\ prescription to deal separately with statistical and systematic errors\label{fig-smdev}} 
\end{center}
\end{figure}

 Most of these observables, either from tree (first two categories) or loop (last three ones) contributions, are established individual benchmarks to constrain or measure the parameter space of the 2HDM Type II or
 cognate supersymmetric models~\cite{2hdmexp,Gfitter}. The measurements of these observables and their experimental uncertainties are displayed in Figure~\ref{fig-smdev} together
 with their SM prediction at $95\%$~CL. Figure~\ref{fig-smdev} shows an overall fair agreement between the various observables and their SM predictions, with the notable exception
 of the tauonic $B^+$ decay $B\rightarrow\tau\nu$ (the deviation is yet lower than $3$ $\sigma$). Let us notice that the recent CLEO measurements~\cite{cleo_update} of
 $D_s\rightarrow\mu\nu$ and $D_s\rightarrow\tau\nu$ decays are now in agreement with their SM predictions in contrast to the situation reported at the time of the 2008 summer
 conferences~\cite{cleo_summer}.

One of the main objectives of the analysis reported in this article consists in determining whether a 2HDM Type II can accommodate the discrepancy coming from the large value of
$B\rightarrow\tau\nu$ as measured by the $B$ factories. A second aim is to determine the allowed region of the parameter space ($m_H$, $\tan \beta$) as constrained by above set of
low-energy observables, corresponding to $\Delta F=1$ tree processes or loop-induced processes featuring a single charged Higgs exchange. This region of parameter space can be
compared to the limits set by LEP from the (absence of) direct production of charged Higgs bosons. 

\section{Observables and theoretical context}

This section details the theoretical predictions for our set of observables in the SM and how they are modified in the context of charged Higgs contributions. A summary of the
relevant measurements and parameters together with their corresponding uncertainties is given in Tables~\ref{tab-inputs-obs} and \ref{tab-inputs-par}. 

\begin{table}
\begin{center}
\begin{tabular}{c|cccc}
  \hline
  \hline
  Input & Value & Unit & Accuracy & Reference \\ 
  \hline
  \hline
  \multicolumn{5}{c}{\bf Branching Ratios } \\
  \hline
  \hline
  $\Gamma[K\rightarrow \mu \nu]/\Gamma[\pi\rightarrow \mu \nu]$                        & $1.336 \pm 0.003$         &            & $(0.2\%)$         & \cite{FlaviaNet} \\
  $\Gamma[K^0 \rightarrow \pi \mu \nu]/\Gamma[K^0 \rightarrow e \nu]$                  & $0.6640 \pm 0.0026$       &            & $(0.4\%)$         & \cite{FlaviaNet} \\
  $\Gamma[\tau \rightarrow K \nu]/\Gamma[\tau \rightarrow \pi \nu]$                    & $6.370 \pm 0.215$         & $10^{-2}$  & $(3.4\%)$         & \cite{PDG} \\
  ${\cal B}[D\rightarrow \mu \nu]$                                                     & $3.82 \pm 0.32 \pm 0.09$  & $10^{-4}$  & $(8.4\%, 2.4\%)$  & \cite{CLEO} \\
  ${\cal B}[D_s\rightarrow \mu \nu]$                                                   & $5.93 \pm 0.40$           & $10^{-3}$  & $(6.7\%)$         & \cite{Stone}\cite{cleo_update} \\
  ${\cal B}[D_s\rightarrow \tau \nu]$                                                  & $5.62 \pm 0.44$           & $10^{-2}$  & $(7.8\%)$         & \cite{Stone}\cite{cleo_update} \\
  ${\cal B}[B\rightarrow \tau \nu]$                                                    & $1.73 \pm 0.35$           & $10^{-4}$  & $(20\%)$          & \cite{BaBar},\cite{Belle} \\
  ${\cal B}[B\rightarrow D \tau \nu]/{\cal B}[B\rightarrow D \ell \nu]$                & $0.416 \pm 0.128$         &            & $(31\%)$          & \cite{bdtaunubelle,bdtaunubabar} \\
  $\Delta m_d$                                                                         & $0.507 \pm 0.005$         & $ps^{-1}$  & $(1.0\%)$         & \cite{PDG} \\
  $\Delta m_s$                                                                         & $17.77 \pm 0.12$          & $ps^{-1}$  & $(0.7\%)$         & \cite{CDF-Dms} \\
  $\Gamma[Z\rightarrow b \bar{b}]/\Gamma[Z\rightarrow {\rm hadrons}]$                  & $0.21629 \pm 0.00066$     &            & $(0.3\%)$         & \cite{EWFit} \\
  ${\cal B}[\bar{B}\rightarrow X_s\gamma]/{\cal B}[\bar{B}\rightarrow X_ce\bar{\nu}]$  & $3.346 \pm 0.251$         & $10^{-3}$  & $(7.5\%)$         & \cite{HFAGbsg} \\
\hline
  \hline
\end{tabular}
\caption{\small \it Branching ratios used as inputs for the global 2HDM Type II analysis. They are listed and their values are given with their
absolute uncertainty, their relative accuracy and the reference from where the value was taken. When two uncertainties are given, the first one is statistical and the second
is systematic (often from theoretical origin).
\label{tab-inputs-obs}}
\end{center}
\end{table}

\begin{table}
\begin{center}
\begin{tabular}{c|cccc}
  \hline
  \hline
  Input & Value & Unit & Accuracy & Reference \\ 

  \hline
  \hline
  \multicolumn{5}{c}{\bf Decay Constants} \\
  \hline
  \hline
  $f_K/f_{\pi}$      & $1.205 \pm 0.0012 \pm 0.0095$  &      & $(0.1\%,0.8\%)$    & \cite{lqcd_ckmfitter} \\
  $f_{D_s}/f_{D_d}$  & $1.186 \pm 0.0048 \pm 0.0010$ &       & $(0.4\%,0.1\%)$    & \cite{lqcd_ckmfitter} \\
  $f_{D_s}$          & $246.3 \pm 1.2 \pm 5.3$        & MeV  & $(0.5\%,2.2\%)$    & \cite{lqcd_ckmfitter} \\
  $f_{B_s}/f_{B_d}$  & $1.199 \pm 0.008 \pm 0.023$    &      & $(0.7\%,1.9\%)$    & \cite{lqcd_ckmfitter} \\
  $f_{B_s}$          & $228 \pm 3 \pm 17$             & MeV  & $(1.3\%,7.5\%)$    & \cite{lqcd_ckmfitter} \\ 
  \hline
  \hline
  \multicolumn{5}{c}{\bf Semileptonic Form Factors} \\
  \hline
  \hline
  $\rho^2$  & $1.19 \pm 0.04 \pm 0.04$       &     & $(3.3\%,3.3\%)$  & \cite{HFAGbsg} \\
  $\Delta$  & $0.46 \pm 0 \pm 0.01$          &     & $(0, 2.2\%)$     & \cite{MesciaKamenik} \\
  $M_V$     & $878 \pm 6$                    & MeV & $(0.6\%)$        & \cite{PDG} \\
  $f_+(0)$  & $0.9653 \pm 0.0028 \pm 0.0048$ &     & $(0.3\%, 0.5\%)$ & \cite{lqcd_ckmfitter} \\
  \hline
  \hline
  \multicolumn{5}{c}{\bf $B\bar{B}$ mixing} \\
  \hline
  \hline
  $\hat{B}_{B_s}$                  & $ 1.28 \pm 0.02 \pm 0.03$   &  & $(1.6\%, 2.3\%)$  &  \cite{lqcd_ckmfitter} \\
  $\hat{B}_{B_s} / \hat{B}_{B_d}$  & $ 1.05 \pm 0.01 \pm 0.03$   &  & $(1.0\%, 2.9\%)$  &  \cite{lqcd_ckmfitter} \\
  $\eta_B$                         & $ 0.5510 \pm 0 \pm 0.0022$  &  & $(0, 0.4\%)$      &  \cite{Buchalla,Lenz_etaB} \\
  \hline
  \hline
  \multicolumn{5}{c}{\bf $Z \rightarrow b\bar{b}$} \\
  \hline
  \hline
  $\Delta\alpha_{had}^{(5)}[m_Z]$  & $ 0.02758 \pm 0.00035$   &  & $(1.3\%)$  &  \cite{EWFit} \\
  \hline
  \hline
  \multicolumn{5}{c}{\bf $b\rightarrow s \gamma$ parameterization } \\
  \hline
  \hline
  $C$             & $0.546 \pm 0 \pm 0.033$   &     & $(0, 6.0\%)$  & \cite{bsgammaNewC} \\
  $m_t^{pole}$    & $172.4 \pm 1.2$     & GeV       & $(1.2\%)$     & \cite{TevatronEWG} \\
  $\alpha_s(m_Z)$ & $0.1176 \pm 0.0020$ &           & $(1.7\%)$     & \cite{PDG} \\
  \hline
  \hline
  \multicolumn{5}{c}{\bf Running Quark Masses } \\
  \hline
  \hline
  $\overline{m}_u(2 \; {\rm GeV})$            & $2.40 \pm 0 \pm 0.90$               & MeV & $(0, 38\%)$          & \cite{PDG} \\
  $\overline{m}_d(2 \; {\rm GeV})$            & $4.75 \pm 0 \pm 1.25$               & MeV & $(0, 26\%)$          & \cite{PDG} \\
  $\overline{m}_s(2 \; {\rm GeV})$            & $96 \pm 0 \pm 30$                   & MeV & $(0, 31\%)$          & \cite{PDG} \\
  $\overline{m}_c(m_c)$                       & $1.286 \pm 0.013 \pm 0.040$         & GeV & $(1.0\%, 3.1\%)$     & \cite{lqcd_ckmfitter} \\
  $\overline{m}_b(m_b)$                       & $4.243 \pm 0 \pm 0.043$             & GeV & $(0, 1.0\%)$         & \cite{HFAGbsg} \\       
  \hline
  \hline
\end{tabular}
\caption{\small \it Parameters used as inputs for the global 2HDM Type II analysis. The parameters entering the calculations are listed and their values are given with their
absolute uncertainty, their relative accuracy and the reference from where the value was taken. When two uncertainties are given, the first one is statistical and the second
is systematics, often from theoretical error, hence treated in the \rfit\ scheme. For the latter systematics, whenever individual contributions are listed in the quoted reference we
combine them linearly instead of quadratically, to stay consistent with the \rfit\ scheme. Therefore, our systematics errors can be larger than the one given in the corresponding
references. Further note that the scale invariant top quark mass $\overline{m}_t(m_t)$ is computed from $m_t^{pole}$, following eq.~(33) of \cite{RunningQuarkMass} with $n_f = 5$
active flavours. \label{tab-inputs-par}}
\end{center}
\end{table}

\subsection{Leptonic decays}

The decay of a charged meson $M$ into a leptonic pair $\ell \nu_\ell$ is mediated in the SM by 
a charged weak boson, with the branching ratio:  
\begin{align} \label{eq-Mlnu}
{\cal B}[M\rightarrow \ell \nu_\ell]_{\rm SM} = \frac{G_F^2 m_M m_{\ell}^2 }{ 8 \pi }
\left( 1 - \frac{m_{\ell}^2}{m_M^2} \right)^2 |V_{q_uq_d}|^2 f_M^2 \tau_M ( 1 + \delta_{EM}^{M \ell 2} ),
\end{align}		 
where $q_u$ ($q_d$) stands for the up-like (down-like) valence quark of the meson respectively,
$V_{q_uq_d}$ is the relevant CKM matrix element, $f_M$ is the decay constant of the meson $M$ (describing how strong the coupling of the meson to the axial current can be) and
$\tau_M$ its lifetime. The corrective factor $\delta_{EM}^{M \ell 2}$ stands for channel-dependent electromagnetic radiative corrections. In this work, they are taken into account
in the case of the lighter mesons ($\pi$ and $K$), where their impact is estimated to be at the level of $2-3\%$~\cite{Kl2pil2em,FlaviaNet}, and for the $D$ meson, where the effect
is~$1\%$~\cite{Dl2em,cleo_update}. As far as $B$-related observables are concerned, the experiments take into account soft photons corrections derived from their Monte Carlo
simulated data, and we will assume that no further correction is required for these branching ratio (setting $\delta_{EM}^{B \ell 2}=0$). 

\par

The experimental accuracies for the branching ratios are given in the Table~\ref{tab-inputs-obs}, lying within $\simeq 0.2-31\%$ depending on the leptonic decay of interest. The main
theoretical uncertainty arises from the decay constant, which is a non-perturbative quantity to be estimated by theoretical methods, such as quark models, sum rules, or lattice
QCD (LQCD) simulations. We opt for the latter, since they provide well-established methods to compute these observables with a good accuracy and a satisfactory theoretical control.  

Over the last few years, many new estimates of the decay constants have been issued by different lattice collaborations, with different ways to address the errors. A part of the
uncertainties has a clear statistical interpretation: lattice simulations evaluate correlators in an Euclidean metric expressed as path integrals using Monte Carlo methods, whose
accuracy depends crucially on the size of the sample of gauge configurations used for the computation. But systematics are also present and depend on the strategies of computation
chosen by competing lattice collaborations: discretisation methods used to describe gauge fields and fermions on a lattice, parameters of the simulations, such as the size of the
(finite) volumes and lattice spacings used for simulations, the masses of the quarks that can be simulated, and the number of dynamical flavours (2 and 2+1 being the most
frequent). In relation with these choices, the extrapolation of the results to physical parameters can be subject to different theoretical treatments (chiral perturbation theory,
heavy-quark expansion\ldots), going beyond a naive linear extrapolation. 

The combination of lattice values with different approaches to address the error budget is a critical point of most global analyses of the flavour physics data, even though
the concept of the theoretical uncertainty for such quantities is ill-defined (and hence is the combination of them). The CKMfitter group has collected the relevant LQCD
calculations of the decay constants $f_{B_d}$, $f_{B_s}$, $f_{D_s}$, $f_{D}$, $f_{K}$, $f_{\pi}$ (as well as bag factors ${\cal B}_{B_d}$, ${\cal B}_{B_s}$ and ${\cal B}_{K}$) and
designed an averaging method aiming at providing a systematic, reproducible and to some extent conservative scheme~\cite{lqcd_ckmfitter}. These lattice averages are the input
parameters used in the fits presented in this paper. 

\par

In the specific case of light mesons (kaons and pions), the ratio of the decay constants $f_K/f_{\pi}$ is significantly better determined than the individual decay constants. It
is hence worth considering the ratio $K_{\ell 2}/\pi_{\ell 2}$ of the kaon and pion leptonic partial widths instead of the individual branching ratios. It explicitly writes in the
SM as~\cite{FlaviaNet}: 
\begin{align} \label{eq-Kl2Opil2}
\frac{\Gamma[K\rightarrow\mu\nu]_{\rm SM}}{\Gamma[\pi\rightarrow\mu\nu]_{\rm SM}} = \frac{m_K}{m_\pi}
\left( \frac{1 - m_l^2/m_K^2}{1 - m_l^2/m_\pi^2} \right)^2 \left|\frac{V_{us}}{V_{ud}}\right|^2 \left(\frac{f_K}{f_\pi}\right)^2 ( 1 + \delta_{EM}^{K\ell2/\pi \ell 2} ).
\end{align}
Let us notice that short-distance radiative corrections cancel out in the ratio. The long-distance corrections are accounted for by the parameter $\delta_{EM}^{K \ell 2/\pi \ell
2} = -0.0070 \pm 0.0035$, which can be computed using chiral perturbation theory~\cite{Kl2pil2em}. Similarly we also consider the ratio of taus to kaons to taus to pions decays.
The latter writes: 
\begin{align} \label{eq-tauK2Otaupi2}
\frac{\Gamma[\tau\rightarrow K\nu]_{\rm SM}}{\Gamma[\tau\rightarrow\pi\nu]_{\rm SM}} =
\left( \frac{1 - m_K^2/m_\tau^2}{1 - m_\pi^2/m_\tau^2} \right)^2 \left|\frac{V_{us}}{V_{ud}}\right|^2 \left(\frac{f_K}{f_\pi}\right)^2 ( 1 + \delta_{EM}^{\tau K 2/\tau \pi 2} ).
\end{align}
The radiative correction is taken from~\cite{taupiK} and reads $\delta_{EM}^{\tau K 2/\tau \pi 2} = 0.0003$. The uncertainty is coming from $\delta_{EM}^{K \ell 2/\pi\ell 2}$ and
$\delta_{EM}^{(\tau K 2/K \ell 2)/(\tau \pi 2/\pi\ell 2)}=0.0073 \pm 0.0027$, both treated as \rfit\ uncertainties. 

\par 
On the experimental side, the latest measurements of $B\rightarrow \tau \nu$ branching ratio from the $B$ factories \cite{BaBar,Belle} have been taken into account. The two
experiments BaBar and Belle find results in fairly good agreement and the weighted average of ${\cal B}[B \rightarrow \tau \nu]$ exhibits a departure from the SM prediction from
the global CKM fit (more than $2 \sigma $), due to tension between $\sin(2 \beta)$ and ${\cal B}[B \rightarrow \tau \nu]$, as pointed out in~ref.~\cite{cleo_summer}.

\par 

\begin{figure}
\begin{center}
\begin{picture}(40,30)(0,0)
\SetScale{1}
\Oval(10,50)(30,10)(0)
\ArrowLine(10,20)(50,50)
\ArrowLine(50,50)(10,80)
\Photon(50,50)(80,50){4}{4}
\ArrowLine(100,30)(80,50)
\ArrowLine(80,50)(100,70)
\Text(33,28)[c]{$\ell^-$}
\Text(33,7)[c]{$\bar\nu_\ell$}
\Text(10,28)[c]{$\bar{q}_d$}
\Text(10,7)[c]{$q_u$}
\Text(25,22)[c]{$W^+$}
\Text(3,18)[c]{$M$}
\end{picture} 
\qquad
\begin{picture}(40,30)(0,0)
\SetScale{1}
\Oval(10,50)(30,10)(0)
\ArrowLine(10,20)(50,50)
\ArrowLine(50,50)(10,80)
\DashLine(50,50)(80,50){4}
\ArrowLine(100,30)(80,50)
\ArrowLine(80,50)(100,70)
\Text(33,28)[c]{$\ell^-$}
\Text(33,7)[c]{$\bar\nu_\ell$}
\Text(10,28)[c]{$\bar{q}_d$}
\Text(10,7)[c]{$q_u$}
\Text(24,22)[c]{$H^+$}
\Text(3,18)[c]{$M$}
\end{picture} 
\end{center}

\vspace{-0.7cm}

\caption{\it \small Leptonic decay of a meson through the exchange of a $W$ boson (left) and a charged Higgs (right).\label{fig:leptonic}}
\end{figure}
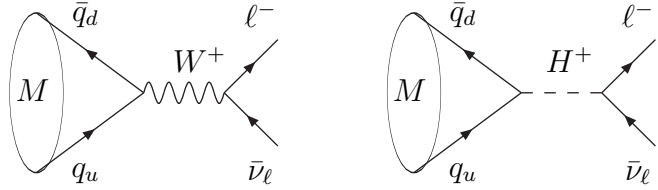

In two Higgs doublet models, purely leptonic decays receive an additional contribution from charged Higgs, as illustrated in Figure~\ref{fig:leptonic}. It turns out that in this
case, this correction can be factorized from the SM prediction~\cite{2hdmref4,2hdmexp,2hdm2ds}: 
\begin{align} \label{eq-MlnuHchIntro}
{\cal B}[M\rightarrow \ell \nu] = {\cal B}[M\rightarrow \ell \nu]_{\rm SM}( 1 + r_H )^2,
\end{align}			 
where the corrective factor $r_H$ writes in 2HDM Type II:
\begin{align} \label{eq-MlnuHch}
r_H = \left(\frac{m_{q_u}-m_{q_d} \tan^2 \beta}{m_{q_u}+m_{q_d}}\right)\left(\frac{m_M}{m_{H^+}}\right)^2. 
\end{align}				 
A comment is in order concerning the structure of eq.~(\ref{eq-MlnuHch}). Let us suppose that we have a perfect agreement between the measurement of a purely leptonic decay and
its SM expression. There are actually in this case two distinct solutions in the 2HDM Type II framework: 
\begin{itemize}
\item $r_H=0$, which can be obtained easily by sending $m_{H^+}$ to infinity. This \emph{decoupling solution} corresponds to a general way of recovering SM predictions by assuming
that all additional particles are very massive. 
\item $r_H=-2$, which corresponds to a linear correlation between $m_{H^+}$ and $\tan\beta$. This \emph{fine-tuned solution} depends on the mass of the meson and those of its
valence quarks, and it is thus different from one meson to another. 
\end{itemize}

More generally, if we have a good agreement between the SM prediction and the measurement (as is the case for most leptonic decays), the 2HDM Type II fit will favour two regions:
one region of high Higgs mass (related to the decoupling solution) and one diagonal band (in relation with the fined-tuned solution) in the parameter space ($m_{H^+}$, $\tan\beta$). 

\subsection{Semileptonic $B$ decay: $B\rightarrow D \tau \nu$}

Purely leptonic decays of mesons intertwine electroweak and strong interactions. However, the role of strong interaction boils down to the presence of a decay constant, to be
assessed through theoretical methods. Semileptonic decays are more complicate to describe, since they involve form factors with a non-trivial dependence on the transfer momentum.

If the form factors are known with a sufficient accuracy, semileptonic branching ratios start becoming valuable constraints on New Physics models -- for instance, the comparison
between leptonic and semileptonic decays provides a good test of the $V-A$ structure of weak interactions. The BaBar and Belle experiments recently published first measurements of
${\cal B}(B\rightarrow D\tau\nu)$ \cite{bdtaunubelle,bdtaunubabar}.  

An interesting observable is the normalized branching ratio ${\cal R}_{B\rightarrow D\tau\nu}={\cal B}[B\rightarrow D\tau\nu]/{\cal B}[B\rightarrow D e \nu]$, which corresponds to
a $b \to c$ transition, with a CKM factor much larger than the purely leptonic $B$ decay (and thus easier to study experimentally). 
In principle, the relevant form factors can be studied using LQCD simulations, as long as one is interested in a limited region of space-like transfer momentum, where both
incoming and outgoing states consist of a single meson and no final state-interaction occurs. In order to extend the range of determination of form factors, fits combining lattice
information and $B\rightarrow D \ell \nu$ can be performed~\cite{MesciaKamenik}, in order to constrain the shape of the relevant vector and scalar form factors over the whole
kinematic regime.  

In the case of 2HDM-II models, the scalar form factor is a key ingredient, since it encodes the impact of the charged Higgs exchange in the semileptonic decay. Unfortunately,
scalar form factors are notoriously difficult to handle on the lattice, and require dedicated methods and significant computing power to be estimated~\cite{latticebtod}. Due to
helicity suppression, this scalar contribution arises with a factor $m_\ell^2/m_B^2$ in the amplitude, which means that only $B\to D\tau\nu_\tau$ is sensitive to this
contribution~\footnote{In principle, a similar analysis could be applied for $B\to D^*\tau\nu$. However it involves four form factors which are poorly known and out of which only
one would be sensitive to charged Higgs exchange~\cite{BtoDstar}.}. 
Following ref.~\cite{MesciaKamenik}, we write the ratio ${\cal R}_{B\rightarrow D\tau\nu}$ as a second order polynomial in the charged Higgs coupling, as:
\begin{align} \label{eq-BDtaunu}
{\cal R}_{B\rightarrow D\tau\nu}= a_0 + a_1 {\cal R}e[ s_{H} ] + a_2 |s_{H}|^2.
\end{align}		    
The factor $s_{H}$ accounting for the charged Higgs coupling was taken as:   
\begin{align} \label{eq-NHch}
s_{H}=-\frac{\tan\beta^2}{1-m_c/m_b} \cdot \frac{m_B^2-m_D^2}{m_{H^+}^2}. 
\end{align}		 
The polynomial coefficients $a_i$ in eq.~(\ref{eq-BDtaunu}) depend on the slope parameter  of the vector form factor $\rho^2$ and the scalar form factor $\Delta$ (taken as a
constant) describing the $B \rightarrow D$ vector and scalar form factors. A detailed statistical treatment of the theoretical uncertainties {\it \`a la} \rfit\ requires an
explicit knowledge of these dependencies. Therefore we do not use the final result of ref.~\cite{MesciaKamenik} but we rather integrate their expression for the $B\rightarrow D
\tau \nu$ differential branching ratios for various values of the parameters governing the form factors. We checked that the variations of the branching ratio with $\rho^2$ and
$\Delta$ are smooth over the range of uncertainty of these latter parameters. Consequently, the coefficients $a_i$ are well parameterized as: 
\begin{align} \label{eq-BDParam}
a_0 = 0.2970 + 0.1286 \cdot d\rho^2 + 0.7379 \cdot d\Delta, \\
a_1 = 0.1065 + 0.0546 \cdot d\rho^2 + 0.4631 \cdot d\Delta, \\
a_2 = 0.0178 + 0.0010 \cdot d\rho^2 + 0.0077 \cdot d\Delta,
\end{align}
where we have averaged over $B_{d,u} \rightarrow D_{d,u}$ modes and where $d\rho^2 = \rho^2-1.19$ and $d\Delta = \Delta-0.46$ are the variations of the $\rho^2$ and $\Delta$
parameters. Let us mention that we have also investigated the constraints derived in ref.~\cite{NiersteBDtaunu} from the same observables, and we checked that we obtained very comparable results,
even though the theoretical expressions are quite different. 

\subsection{Semileptonic kaon decay: $K\rightarrow \pi \ell \nu$}

The process $K\rightarrow \pi \ell \nu$ is known to very good experimental and theoretical accuracies (better than $1\%$) and hence is included in this analysis, even though it
involves lighter fermions. The branching ratio ${\cal B}[K^0\rightarrow \pi \mu \nu]/{\cal B}[K^0\rightarrow \pi e \nu]$ writes: 
\begin{align} \label{eq-KpiGlobal}
\frac{{\cal B}[K^0\rightarrow\pi\mu\nu]}{{\cal B}[K^0\rightarrow\pi e\nu]}= \frac{\hat{I}_K^\mu}{\hat{I}_K^e}( 1 + 2\delta_{EM}^{K,\mu} - 2\delta_{EM}^{K,e} ),  
\end{align}
where $\delta_{EM}^{K,e} = (5.7 \pm 1.5)\cdot 10^{-3}$ and $\delta_{EM}^{K,\mu} = (8.0 \pm 1.5)\cdot 10^{-3}$ are radiative electromagnetic corrections estimated in Chiral
Perturbation Theory in ref.~\cite{Kl3em} and recalled in  ref.~\cite{FlaviaNet}. The phase space integrals $\hat{I}_K^l$ depend on both the scalar and vector form factors
describing the $K\to\pi$ transitions. In the low energy region of interest here, the vector form factor can be described accurately through resonance saturation (involving the
$K^*$ pole). The scalar form factor is more delicate to describe, but it can be expressed through a dispersion relation involving data on $\pi K$ scattering.  

Exploiting these form factors, a parametrization of the ratio of the phase space integrals is derived in Appendix A:
\begin{align} \label{eq-KpiParam}
\frac{\hat{I}_K^\mu}{\hat{I}_K^e} & = p_I[z]( 1+ p_S[z]\epsilon_G + p_V[z]\Delta M_V ), \\
p_I[z] & = 7.96407 \cdot 10^{-3} z^3 + 2.59205 \cdot 10^{-2} z^2 + 7.82087 \cdot 10^{-2} z + 0.647932, \\
p_S[z] & = 6.04408 \cdot 10^{-5} z^2 + 2.30011 \cdot 10^{-4} z + 4.67096 \cdot 10^{-4} z, \\
p_V[z] & = 4.24716 \cdot 10^{-5} z - 2.69983 \cdot 10^{-5},
\end{align}
where $\Delta M_V=(M_V-892)$ is a parameter related to the mass of the $K^*$ resonance expressed in MeV. The factor $\epsilon_G$ is a free parameter in $[-1;1]$ reflecting
theoretical uncertainties on the scalar form factor. Charged Higgs contributions are accounted for through the parameter $z$ as~\cite{FlaviaNet}: 
\begin{align} \label{eq-KpiZFactor}
z    & = \frac{f_K}{f_\pi}\frac{1}{f_+(0)}+\Delta_{CT} - \frac{m_s \tan^2\beta-m_u}{m_s-m_u}\left(\frac{m_{K^0}^2-m_{\pi^+}^2}{m_{H^+}^2}\right),
\end{align}
where $f_+(0)$ denotes the normalization of the vector form factor at $q²=0 \; {\rm GeV}^2$ and $\Delta_{CT} \in [-0.5;0] \cdot 10^{-3}$, estimated in Chiral Perturbation Theory,
describes the deviation of the scalar form factor from $f_K/f_\pi$  at the Callan-Treiman point. 

\subsection{Radiative decays: $b\rightarrow s \gamma$}

The FCNC decay $b\rightarrow s \gamma$ proceeding through penguin diagrams is a powerful benchmark to constrain the charged Higgs sector of New Physics models. The calculation of
the $\bar{B}\rightarrow X_s\gamma$ branching ratio has been completed up to Next-to-Next Leading Order (NNLO)~\cite{bsgammaNNLO} (see
refs.~\cite{bsgammaEarly}-\cite{bsgammaHchlimit} for details).  
\par
Its starting point is the effective Hamiltonian derived by integrating out the degrees of freedom heavier than the $b$ quark, and expressed as products of effective operators
(describing long-distance effects) with Wilson coefficients (including short-distance effects). The inclusive decay rate is expressed as the 
imaginary part of the relevant correlator of two $b\to s\gamma$ currents, 
which can be expanded in powers of $1/m_b$. Most
of the effort in the field has been devoted to the computation of the perturbative part of the leading contribution to this expansion.

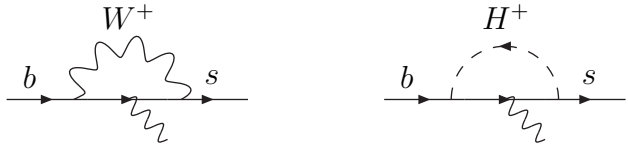
\begin{figure}
\begin{center}
\begin{picture}(40,25)(0,-10)
\SetScale{1}
\ArrowLine(0,0)(30,0)
\ArrowLine(30,0)(60,0)
\ArrowLine(60,0)(90,0)
\PhotonArc(45,0)(20,0,180){4}{5}
\Photon(45,0)(60,-15){3}{3}
\Text(3,3)[c]{$b$}
\Text(27,3)[c]{$s$}
\Text(16,11)[c]{$W^+$}
\end{picture} 
\qquad
\begin{picture}(40,25)(0,-10)
\SetScale{1}
\ArrowLine(0,0)(30,0)
\ArrowLine(30,0)(60,0)
\ArrowLine(60,0)(90,0)
\DashArrowArc(45,0)(20,0,180){4}
\Photon(45,0)(60,-15){3}{3}
\Text(3,3)[c]{$b$}
\Text(27,3)[c]{$s$}
\Text(16,11)[c]{$H^+$}
\end{picture} 
\end{center}

\vspace{-0.7cm}

\caption{\it \small An example of $b\to s\gamma$ transition through $W$ exchange (left) and charged Higgs exchange (right). A second diagram can be drawn with the photon emitted
from the charged boson. 
\label{Fig:BtoSgamma}}
\end{figure}

The NNLO expression for the branching ratio is complicated; for practical purposes, we have chosen to parametrize the results given by the public package SusyBSG~\cite{SusyBSG} (other programs do exist, see for instance~\cite{bsgammanazila}), based on Leading Logarithm (LL) expressions. The SusyBSG code includes NLO perturbative corrections for different theoretical frameworks: SM, 2HDM and MSSM. 

In the 2HDM Type II, the exchange of charged Higgs bosons in addition to charged weak gauge bosons, shown in Figure~\ref{Fig:BtoSgamma}, provides a further contribution to the
relevant Wilson coefficients of the effective Hamiltonian. 
The parameters of the SM prediction, limited to NLO, have been tuned in order to recover the most accurate NNLO result given in ref.~\cite{bsgammaNNLO}. Following the notation in
ref.~\cite{bsgammaQuark}, the normalized branching ratio for $\bar{B}\rightarrow X_s\gamma$ writes:     
\begin{align} \label{eq-bsgamma}
{\cal R}_{b\rightarrow s \gamma} = \frac{{\cal B}[\bar{B}\rightarrow X_s\gamma]}{{\cal B}[\bar{B}\rightarrow X_c\ell \bar{\nu}]} =
\left|\frac{V_{ts}^{*}V_{tb}}{V_{cb}}\right|^2\frac{6\alpha_{\rm EM}}{\pi C}(P+N), 
\end{align}			  
where $P$ denotes the leading contribution in the $1/m_b$ expansion, computed in perturbative QCD, and $N$ the non-perturbative ones, corresponding to higher orders in the $1/m_b$
expansion (starting at $1/m_b^2$). If $P$ can be systematically improved by computing higher and higher orders in perturbation theory, it is hard to provide more than an order of
magnitude for $N$~\cite{bsgammaEarly}. 

The normalization factor $C$ accounts for the phase space difference between charmed semileptonic transition and $\bar{B}\rightarrow X_s\gamma$ decay. For our analysis of 2HDM
Type II, $P$ and $N$ are parametrized using two functions, $A$ and $B$, depending on a reduced set of the relevant input parameters ($\alpha_s(m_Z)$, $m_t^{\rm pole}$ and
$\overline{m}_c(m_c)$).
Making use of the perturbative expression of $P$ at the leading-logarithm level, the functions $A$ and $B$ are defined as: 
\begin{align} \label{eq-bsgammaPara}
P+N = ( C_{7,SM}^{{\rm eff},(0)} + B \Delta C_{7,H^+}^{{\rm eff},(0)} )^2 + A,
\end{align}			  
and fitted to reproduce the results from the SusyBSG package. In eq.~(\ref{eq-bsgammaPara}), the factor $\Delta C_{7,H^+}^{{\rm eff},(0)}$ models the Charged Higgs contributions.
$A$ and $B$, which are independent of the 2HDM Type II parameters $m_{H^+}$ and $ \tan \beta$, exhibit smooth linear variations with the input parameters. Further details on the
parametrization used in this analysis as well as the formulae for $A$ and $B$ functions are given in Appendix B. 

\subsection{Neutral $B$-meson mixing}

In the Standard Model (SM), neutral meson mixing occurs due to box diagrams
with two $W$ exchanges. In the case of $B_d$ and $B_s$ mesons, 
the hierarchical structure of the CKM matrix and the large mass of the top
means that the mixing is dominated by short-distance physics coming from diagrams where the internal fermion lines are top quarks.
In two-Higgs doublet models, the observables related to neutral-meson mixing receive charged Higgs
contributions~\cite{2hdmref1,2hdmref2,2hdmref3,elkaffas}. Indeed,
one gets further diagrams obtained
by replacing one or two $W$ lines by a charged Higgs, yielding~\cite{BBbarMixing}:
\begin{alignat}{2} \label{eq-mixing}
\Delta m_q &  = \frac{G_F^2}{24\pi^2} (V_{tq}V_{tb}^*)^2 \eta_B m_B m_t^2 f^2_{B_q} \hat{B}_{B_q} ( S_{WW} + S_{WH} + S_{HH} ), \\
S_{WW} & = \left( 1 + \frac{9}{1 - x_{tW}} - \frac{6}{(1-x_{tW})^2} - \frac{6 x_{tW}^2 \ln(x_{tW})}{(1 - x_{tW})^3} \right), \\
S_{WH} & = \frac{x_{tH}}{\tan^2 \beta} \left( \frac{ ( 2 x_{HW} - 8 )\ln(x_{tH})}{(1 - x_{HW}) (1 - x_{tH})^2} + 
         \frac{6 x_{HW} \ln(x_{tW})}{(1 - x_{HW})(1 - x_{tW})^2} - \frac{ 8 - 2 x_{tW}}{(1 - x_{tW})(1 - x_{tH})} \right), \\
S_{HH} & = \frac{x_{tH}}{\tan^4 \beta} \left( \frac{1 + x_{tH}}{(1 - x_{tH})^2} + \frac{ 2 x_{tH} \ln(x_{tH})}{(1 - x_{tH})^3} \right), 
\end{alignat}
with $x_{ij} = m_i^2/m_j^2$. $S_{WW}$, $S_{WH}$ and $S_{HH}$ indicate respectively the internal bosonic lines of the corresponding diagrams with an external light quark $q=d,s$. Analyses including radiative corrections are available~\cite{BBbarMixingNLO}, but the leading-order expressions above are sufficient for the required accuracy in our present purposes.

\subsection{Constraint from electroweak precision data: the $Z\rightarrow b\bar{b}$ vertex}

It is time to make an excursion at the border of flavour physics. The $Z\rightarrow b\bar{b}$ vertex has provided opportunities to search for New Physics contributions, due to the heavy masses involved. In particular, the radiative corrections at the vertex might imply charged Higgs exchanges in addition to the standard $W t b$ couplings.  
The partial width $\Gamma[Z\rightarrow b\bar{b}]$ is subject to sizeable QCD corrections. It is hence relevant to define the ratio of the $Z$ partial widths $R_b =
\Gamma[Z\rightarrow b\bar{b}]/\Gamma[Z\rightarrow {\rm hadrons}]$ for which most QCD corrections are suppressed. $R_b$ has been measured by the LEP experiments with a remarkable accuracy~\cite{EWFit}.   
We will not consider neutral Higgs corrections here. Indeed it has been shown~\cite{HaberLogan} that they contribute significantly for large values of $\tan \beta$ only, where $R_b$ is not a competitive observable with respect to the other observables considered here.  On the other hand, the $R_b$ measurement can yield valuable constraints to exclude regions of the 2HDM Type II parameter space at low $\tan\beta$.  

\par 
Following the work described in~refs.~\cite{HaberLogan, Field, Slavich}, we parameterize $R_b$ as:
\begin{align} \label{eq-Rb}
\frac{1}{R_b} = 1 + \frac{K_b}{(\bar{g}_b^L-\bar{g}_b^R)^2+(\bar{g}_b^L+\bar{g}_b^R)^2}, 
\end{align}
where $\bar{g}_b^{L,R}$ are the couplings of left- and right-chirality $b$ quark to the $Z$ boson; $K_b$ is the sum of the
axial and axial-vector couplings of the quark flavours lighter than the $b$ quark, embeding QCD and QED radiative corrections
remaining in $R_b$ as well as effects from the $b$-quark mass. The 3 latter SM quantities have been parameterized to reproduce 
the predictions of the ZFITTER package ~\cite{ZFITTER}, which depends primarily on the top quark mass, $m_t$. However, one has to take care about the correlations between the top
quark mass and the rest of the electroweak parameters. In particular, the neutral Higgs mass, $m_{H^0}$, is constrained by a dedicated Electroweak fit~\cite{EWFit}, following High-$Q^2$~data fit, but excluding the direct measurements of $R_b$, $A_{FB}^{0,b}$ and $A_b$ from the fit inputs. The dependency of the Electroweak $\chi^2$ on $m_t$ is modelled by a correlation factor between $m_t$ and $m_{H^0}$. The full details of the parametrization are provided in Appendix C.
With the top quark mass quoted in table~\ref{tab-inputs-par}, it yields the SM prediction $R_b = 0.21580(4)$, excluding the direct measurement
of $R_b$ as well as $A_{FB}^{0,b}$ and $A_b$ from the fit~\footnote{Let us note incidentally that this prediction is in excellent agreement with the ZPOLE Fit results
of~\cite{EWFit}, which include the measurements of $R_b$, $A_{FB}^{0,b}$ and $A_b$. This is due to the fact that the direct measurement of $m_t$ coincides with its prefered value
from ZPOLE only inputs, which significantly depends on $R_b$.}.

The charged Higgs contribution induces a redefinition of the coupling constants (ref.~\cite{HaberLogan} corrected in~\cite{Slavich}) according to:   
\begin{alignat}{2} \label{eq-gbLR}
\bar{g}_b^L & = \bar{g}_{b,{\rm SM}}^L + \frac{G_F m_W^2 }{ 8 \sqrt{2} \pi^2}\left(\frac{m_t}{m_W}\frac{1}{\tan \beta}\right)^2 F_z\left[\frac{m^2_t}{m^2_{H^+}}\right],\\ 
\bar{g}_b^R & = \bar{g}_{b,{\rm SM}}^R - \frac{G_F m_W^2 }{ 8 \sqrt{2} \pi^2}\left(\frac{m_b}{m_W}\tan\beta\right)^2 F_z\left[\frac{m^2_t}{m^2_{H^+}}\right]. 
\end{alignat}
The function $F_z$ takes into account two-loops corrections following the work developped in~\cite{Slavich} and a parameterization is provided in Appendix C.

Following ~\cite{Field}, we assume that the oblique corrections due to the second Higgs doublet are negligible, so that the modifications from the
2HDM will mainly affect the vertex corrections, and thus modify b-quark observables (and hence $\Gamma(Z\to b\bar{b})$) before any other observable. We are then still allowed to determine the SM prediction for Rb from the Electroweak fit described above (which does not include any b-related observables), out of which we can deduce the value of Rb in 2HDM using eqs. ~\ref{eq-gbLR}.

\section{Individual constraints in the ($m_{H^{\pm}},\tan \beta$) parameter space}

All the fits reported in this paper are performed within the frequentist statistical framework advertised in ~ref.~\cite{ThePapII}. In all analyses, the CKM matrix parameters are
determined simultaneously with the 2HDM additional parameters. Hence, we recall first how the CKM parameters are measured in the framework of the SM and how their determination is
modified once the 2HDM hypothesis is tested.       

\subsection{Standard Model inputs and parameters}

There are four free parameters of interest describing the CKM matrix ($\lambda$, $A$, $\bar \rho$ and $\bar \eta$) in the extended Wolfenstein parametrization~\cite{ThePapII}:
\begin{align} \label{CKMparam}
\lambda=\frac{|V_{us}|}{\sqrt{{|V_{ud}|}^2+{|V_{us}|}^2}}, \qquad A \lambda^2=\frac{|V_{cb}|}{\sqrt{{|V_{ud}|}^2+{|V_{us}|}^2}}, \qquad {\bar \rho}+i {\bar \eta}=
-\frac{V_{ud}V_{ub}^{*}}{V_{cd}V_{cb}^{*}}.   
\end{align}
The dependence of the other CKM matrix elements on these parameters follows from the unitarity of the CKM matrix. 
This definition ensures unitarity at all order of the development in power of $\lambda$ and warrants the $\bar \rho$ and $\bar \eta$ parameters not to depend on phase conventions.
\par
In the SM, $\lambda$ and $A$ are accurately determined: $\lambda$ is measured from super-allowed nuclear transitions and semileptonic kaon decays and $A$ comes from the inclusive
and exclusive semileptonic $b$-hadron decays with charm. On the other hand, the parameters $\bar \rho$ and $\bar \eta$, being respectively the real and imaginary coordinates of
the unitarity triangle (UT) apex, are less constrained. The fit of the CKM matrix in the SM hypothesis and the metrology of its four parameters make use of several observables:    

\begin{itemize}
\item[-] $|V_{ud}|$, $|V_{us}|$ and $|V_{cb}|$ determine the $\lambda$  and $A$ parameters and fix accordingly the length scale of the UT. 
\item[-] $|V_{ub}|$ (including ${\cal B}(B \to \tau \nu)$), $\Delta m_d$ and $\Delta m_s$ are CP-conserving observables, sensitive to the sides of the UT.
\item[-] $\alpha$, $\gamma$, $\sin 2 \beta$, $\cos 2 \beta$ are CP-violating observables measuring the UT angles from $B$-meson decays whereas $|\epsilon_K|$ assesses CP violation
in kaon mixing.     
\end{itemize}

\subsection{2HDM inputs and parameters}\label{sec:inputs}

Let us move to testing the 2HDM Type II hypothesis. It requires to add the two 2HDM Type II parameters ($m_{H^+}$, $\tan \beta$), but also to modify the set of constraints used in
the global SM fit to fix the CKM matrix. We have therefore to split our observables into those
used to fix the CKM matrix, and those needed to constrain the additional parameters from the 2HDM Type II.

First of all, the observables dealing with neutral-meson mixing proceeding through $\Delta F=2$ transitions will receive charged Higgs
contributions~\cite{2hdmref1,2hdmref2,2hdmref3,elkaffas}, and we can use neither the oscillation frequencies of $B_d$ and $B_s$
mesons ($\Delta m_d$ and $\Delta m_s$ respectively) nor the CP-violating parameter $\epsilon_K$, as inputs for the CKM matrix. Moreover, the UT angles $\alpha$ and $\beta$ cannot be used
independently, since their determination relies on an interference between decay and mixing. However, it turns out that the combination of $\alpha$ and $\beta$ inputs in the fit
constitutes a determination of the angle $\gamma$ in which $\Delta F=2$ contributions cancel~\cite{nirgammaT}.

Moreover, the $\Delta F=1$ processes proceed through a $W$-exchange in the SM, and they are thus affected directly by the presence of a charged Higgs
boson. However, we know that these contributions are proportional to the masses of the quarks and charged leptons involved. At low energies, the exchange of a charged Higgs boson
yields four-quark operators with weights $m_1 \cdot m_2/M_H^2$, where $m_1$ and $m_2$ are the masses of two fermions (quarks or charged leptons) involved in the four-quark
operator but not coupled together with a Higgs, as can be seen from eq.~(\ref{HiggsLag}). 

We expect therefore that only processes involving massive quarks and leptons will be very sensitive to 2HDM Type II contributions, which selects naturally some of the processes
considered above for the determination of ($m_{H^+}$, $\tan \beta$): $\mu$ or $\tau$ leptonic decays of $B$, $D_s$ and $D$, $B\to D$ $\mu$ or $\tau$ semileptonic decays (tree processes), $Z\to b\bar{b}$, $b\to s\gamma$ and neutral $B$ meson mixing (processes with a
top-quark loop). 

On the other hand, the CKM matrix, and thus the apex of the unitarity triangle, can be determined by taking $\Delta F=1$ processes where at most one heavy mass is present. This selects:
\begin{itemize}
\item the determination of $\gamma$ from $\alpha+\beta$
($m_b\cdot m_u/m_H^2$, $m_b\cdot m_d/m_H^2$),
\item the determination of $|V_{cb}|$ from semileptonic $b\to c$ decays
($m_b\cdot m_e/m_H^2$, $m_c\cdot m_e/m_H^2$)~\footnote{We should rigorously have taken solely the electronic semileptonic $b \to c$ (and conversely $b \to u$) decays. Yet, the electron and muon averages we are producing are by far dominated by theoretical uncertainties. On top of that, electronic and muonic extractions are in very good agreement and the split averages are not easily available.},
\item the determination of $|V_{ub}|$ from semileptonic $b\to u$ decays
($m_b \cdot m_e/m_H^2$),
\item the determination of $|V_{ud}|$ from super-allowed $\beta$ decays of nuclei
(no heavy mass involved).
\end{itemize}
The determination of $\gamma$ from $B\to DK$ does not enter this list as it scales like $m_b \cdot m_s/M_H^2$.

\begin{figure}
\begin{center}
\epsfig{file=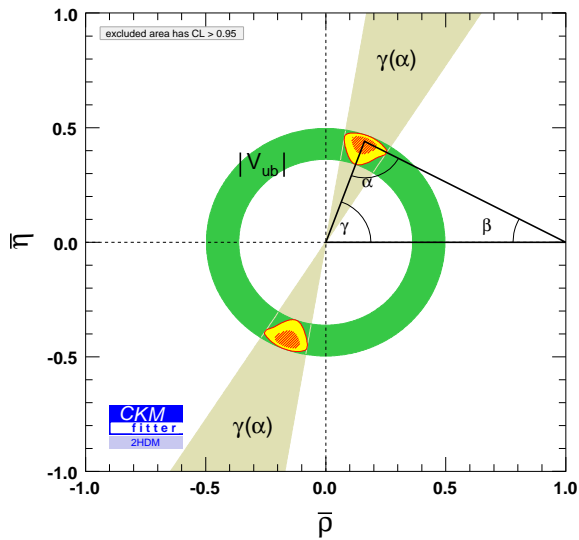,height=80mm,width=80mm}
\caption{\it \small Superimposed individual constraints for the fit comprising the observables which involve light fermions at 95\% CL  (and excluding $\Delta F =2$ observables).
The yellow area is the solution driven by the combination of individual constraints at 95\% CL. The unitarity triangle drawn here is obtained from the global fit displayed in
Figure~\ref{ckm09}. \label{UUT}} 
\end{center}
\end{figure}

 Figure~\ref{UUT} shows the combined constraint in the ($\bar \rho$, $\bar \eta$) parameter space. Though less constraining than the SM global fit, the constraints chosen here
 yield two well-defined symmetrical solutions for the apex of the unitarity triangle. The achieved accuracy is mostly due to the world average $\alpha$ determination, driven by
 the latest $B \rightarrow \rho \rho$ measurements (see ref.~\cite{lqcd_ckmfitter} and references therein).   
                   
Before discussing the combined analysis of all the above constraints, we would like first to focus
on the most stringent individual constraints in the parameter space ($m_{H^+}, \tan \beta$) among the different classes of observables potentially sensitive to charged-Higgs
exchanges. If we impose the Higgs sector to remain in a perturbative regime, an upper bound on the value of $\tan\beta$ can be obtained around 200~\cite{2hdmref3}, and our plots will correspond
to this region (with a logarithmic scale). Two different constraints turn out to constrain the 2HDM Type II very efficiently: the
leptonic decays and the $b\to s\gamma$ branching ratio. 

\subsection{Leptonic and semileptonic decays} 

\begin{figure}[htbp]
\begin{center}

\mbox{\epsfig{file=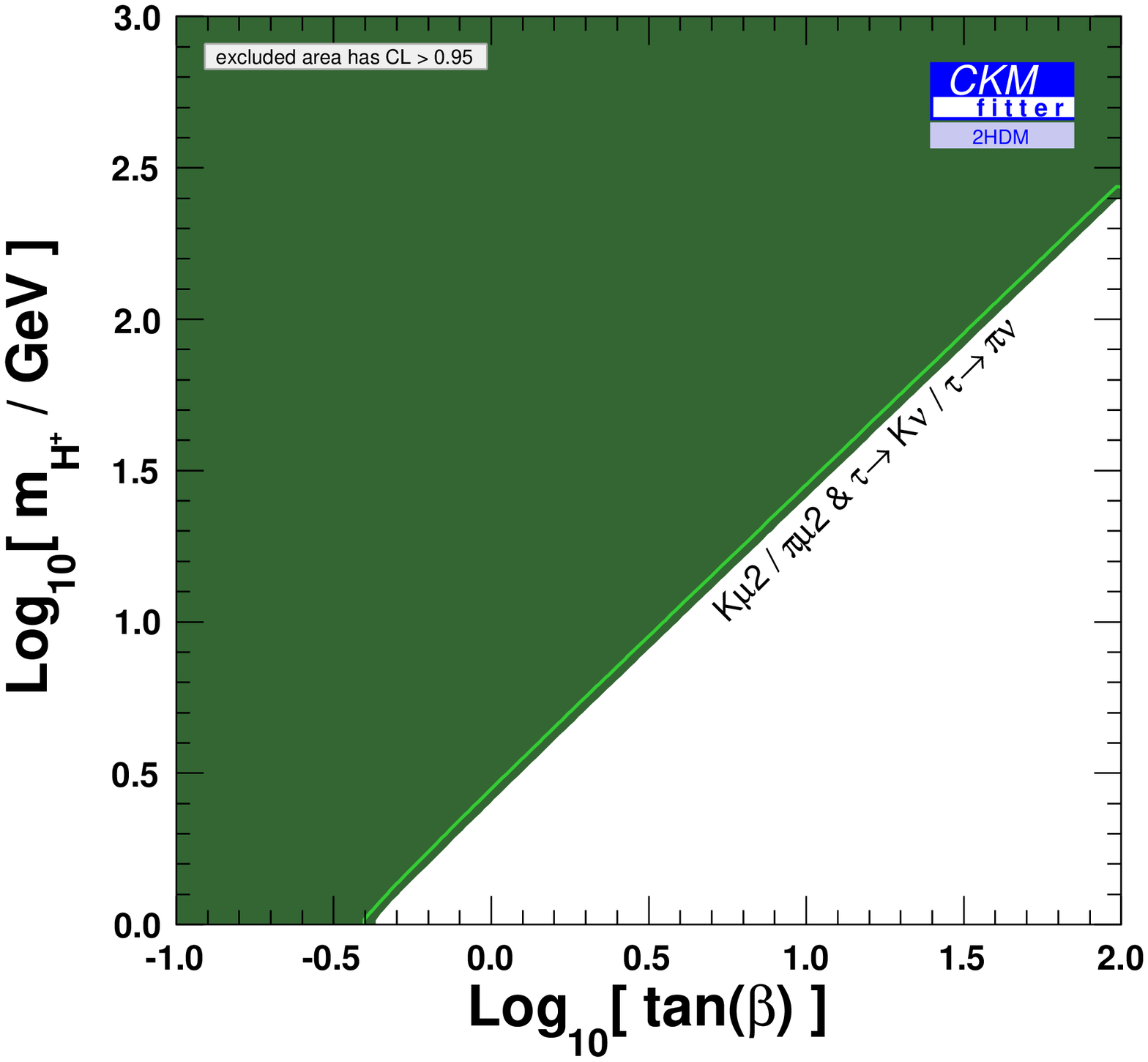,height=60mm,width=60mm}}\mbox{\epsfig{file=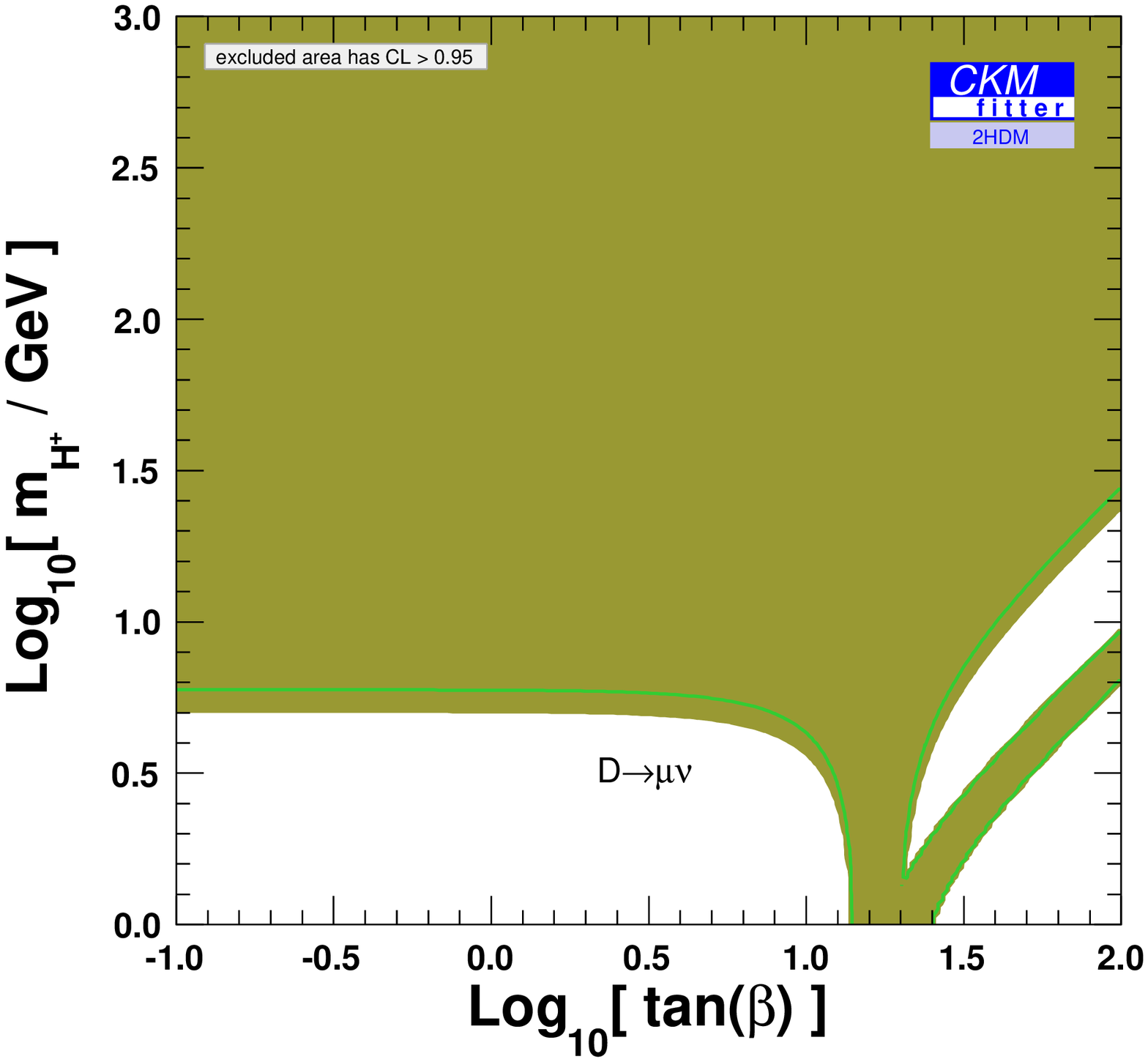,height=60mm,width=60mm}}\\
\mbox{\epsfig{file=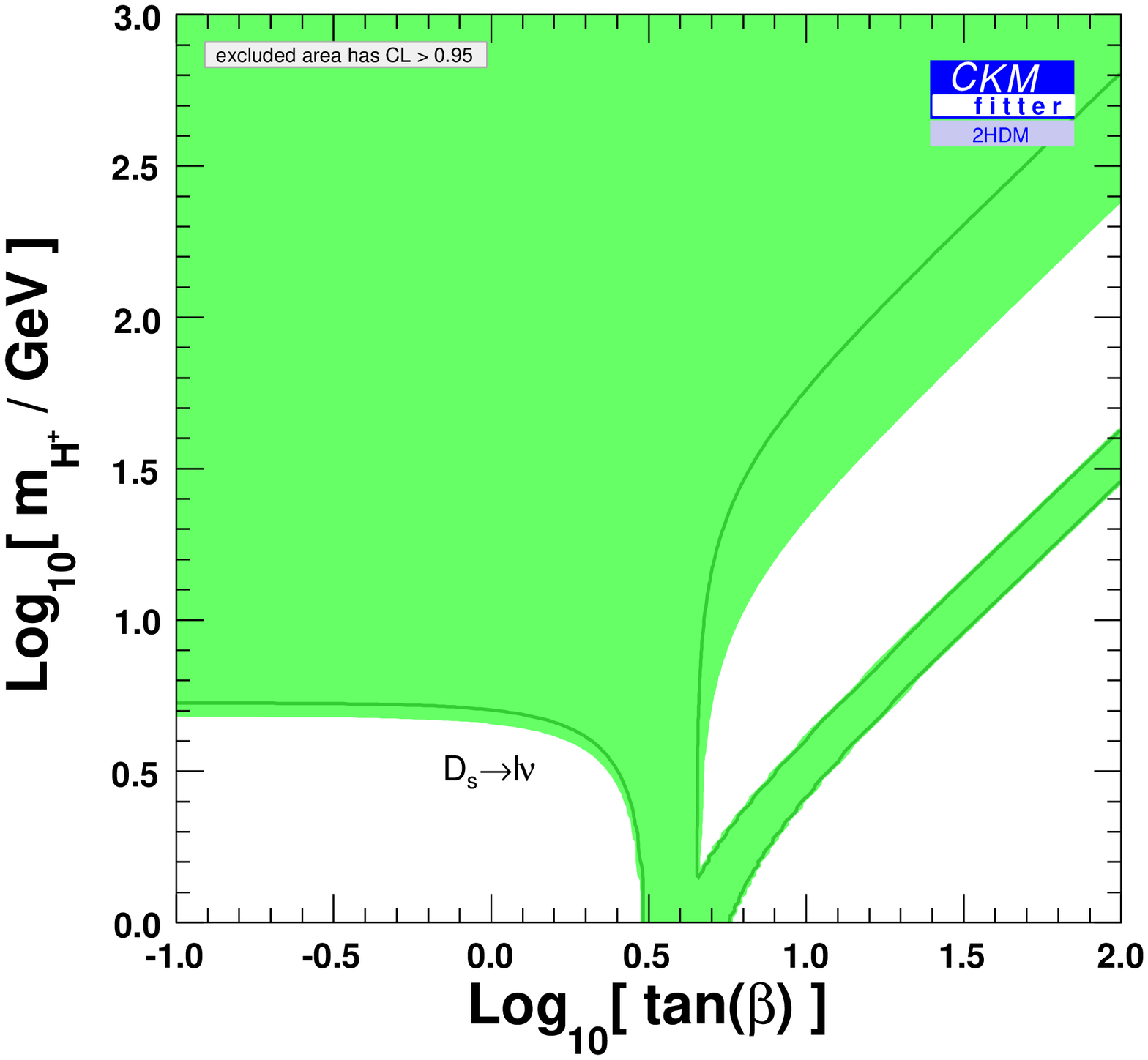,height=60mm,width=60mm}}\mbox{\epsfig{file=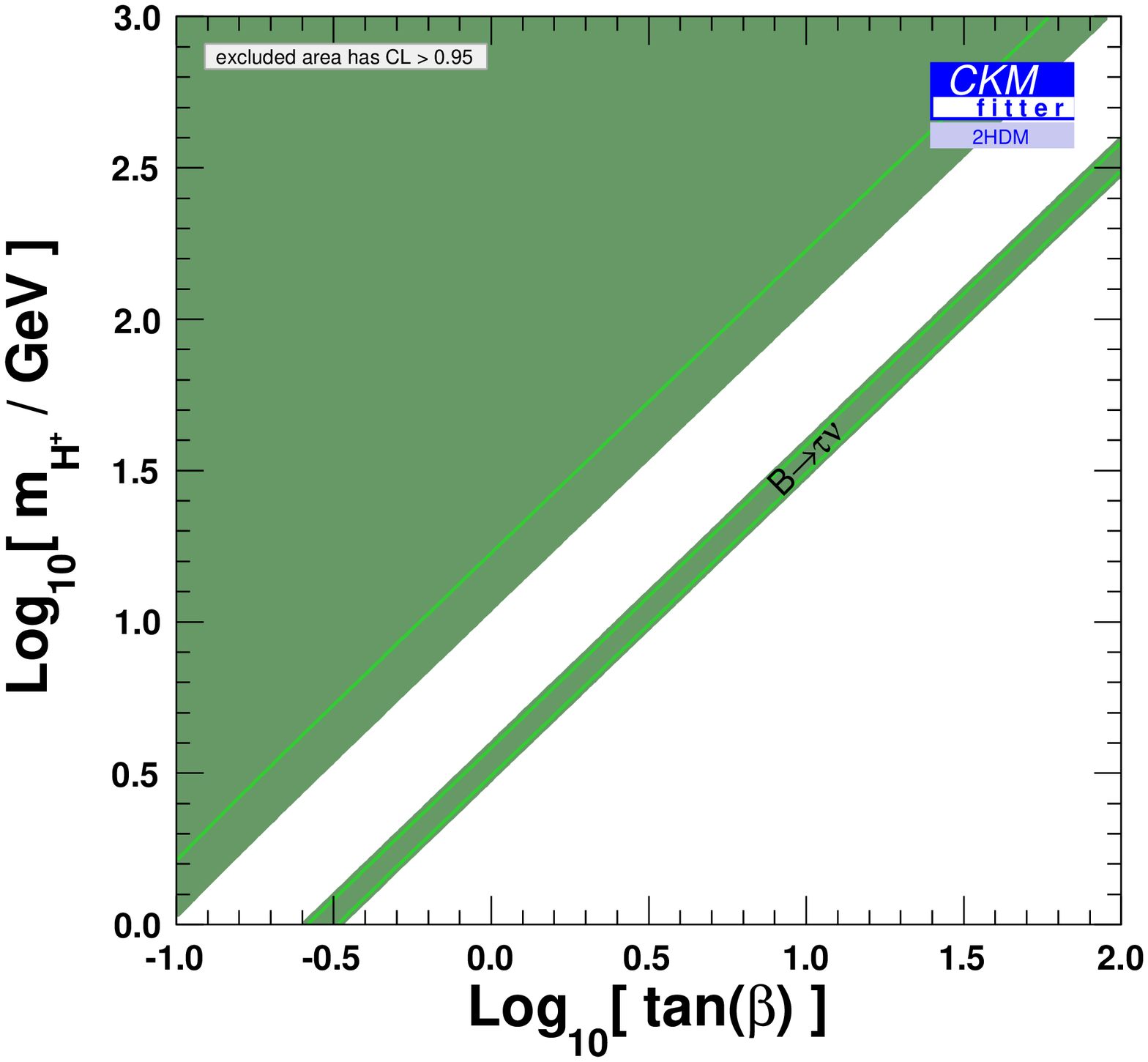,height=60mm,width=60mm}}

\caption{\it \small Constraints on 2HDM parameter space ($m_{H^+}$, $\tan \beta$) from purely leptonic decays. The upper left plot stands for constraints from
$K\to\mu\nu$ and $\tau\to K\nu$ decays, the upper right for $D\to\mu\nu$ decays, the lower left for $D_s$ decays and the lower right for 
$B_d\to\tau\nu$. The superimposed green line delimits the $1\sigma$ confidence area. The excluded regions (white parts of the plots) correspond to more than 95\%CL.\label{fig-leptosplit}}
\end{center}
\end{figure}

\begin{figure}[htbp]
\begin{center}
\mbox{\epsfig{file=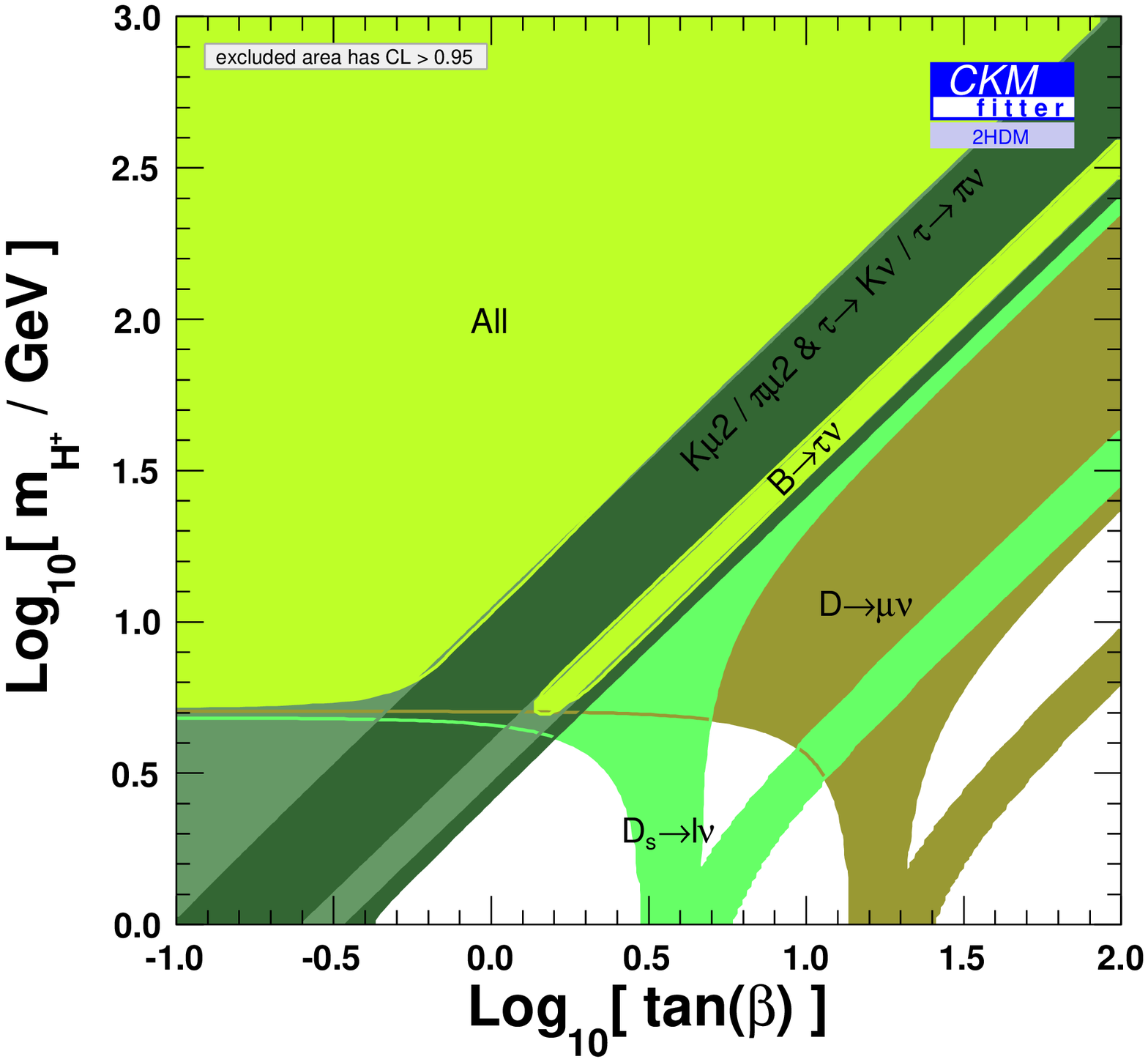,height=80mm,width=80mm}}\mbox{\epsfig{file=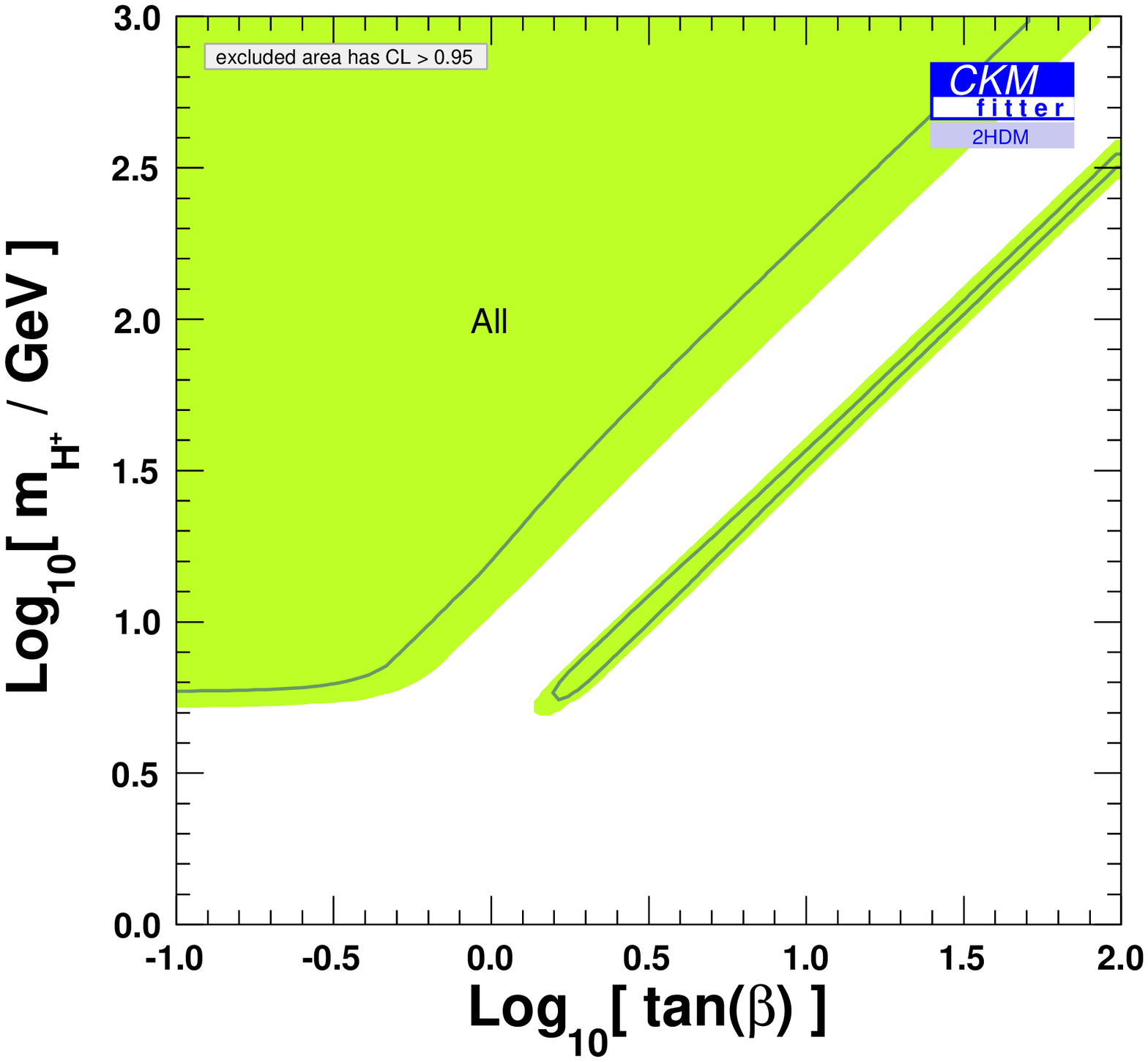,height=80mm,width=80mm}}\\
\caption{\it \small Combined constraints on 2HDM Type II parameter space $(m_{H^+}, \tan \beta)$ from purely leptonic decays of $K$, $D$ and $B$ mesons. The different sets of
observables are defined in the left Figure. Darker to lighter shades of colours correspond to ($K_{\ell 2}/\pi_{\ell 2}+\tau\to K\nu$), ($D\to\mu\nu$), ($B\to\tau\nu$) and
($D_s\to\ell\nu$) individual constraints. The complementary area of the colored region is excluded at $95\%$~CL. For the sake of clarity, the figure on the right displays the
combined constraint alone at $95\%$~CL. The dark green line defines the $1\sigma$ confidence area. 
\label{fig-lepto}}
\end{center}
\end{figure}

As mentioned in the introduction, the only observable which departs significantly from the SM prediction is ${\cal  B}(B^+ \to \tau^+ \nu)$. According to eq.~(\ref{eq-MlnuHch}), 2HDM Type II can accommodate such a sizeably larger value with respect to the SM in only two situations: 
\begin{itemize}
\item at low enough $\tan \beta$ with respect to $\sqrt{m_u/m_b} \simeq 0.02$, to get a positive correction $r_H$ to the SM value,
\item in a fine-tuned scenario (defined in section 2.1) with $r_H < -2$, so that the 2HDM Type II contribution $(1+r_H)^2$ enhances significantly the SM prediction.
\end{itemize}
The $95\%$~CL constraints derived in the plane ($m_{H^+}$, $\tan \beta$) from the various leptonic decays are shown in Figure~\ref{fig-leptosplit}, in a log-log scale. Let us recall that
the charged Higgs contributions are identical for $K_{\mu 2}/\pi_{\mu 2}$ and $\tau \rightarrow K \nu / \tau \rightarrow \pi \nu$. However, since the experimental value of
$K_{\mu 2}/\pi_{\mu 2}$ is yet well better known than $\tau \rightarrow K \nu / \tau \rightarrow \pi \nu$, the latter only has a marginal contribution to the combination.
\par
Combining all constraints from leptonic decays, the minimum $\chi^2$ value of $\chi^2_{min}=14.8$ ($p$-value~$=85.5 \pm 0.3 \%$) is found at small  $m_{H^+}$ where $B
\to \tau \nu$ and $D \to \ell \nu$ fine-tuned 
regions overlap. The very small charged Higgs mass, excluded by direct searches at LEP, reflects the fact that 2HDM Type II can hardly accommodate the large value of
the measured $B \to \tau \nu$ branching ratio at low masses of the charged Higgs but in a fine-tuned scenario, as can be seen in Figure~\ref{fig-lepto}.     

At values of $\tan \beta$ sufficiently large compared to $\sqrt{m_{q_u}/m_{q_d}}$, the $H^{\pm}$ contribution to the branching ratio behaves as $r_H \simeq -(\tan^2 \beta) m_{\rm
eff}^2/m_{H^+}^2$ with $m_{\rm eff} =  m_M \sqrt{m_{q_u}/(m_{q_u}+m_{q_d})}$. Since the corrections to the SM branching ratio are at least quadratic in the effective mass $m_{\rm
eff}$, the $B\rightarrow \tau \nu$ branching ratio sets a constraint on the high-$m_{H^+}$ region. Let us mention that the ratio $K_{\mu 2}/\pi_{\mu 2}$, accurately measured and
predicted, plays a significant role in the combined leptonic limits. As shown in Figure~\ref{fig-lepto}, this additional observable disfavours the $B\rightarrow \tau \nu$
fine-tuned band in the region $\tan \beta \simgt 10$.

\begin{figure}[htbp]
\begin{center}
\epsfig{file=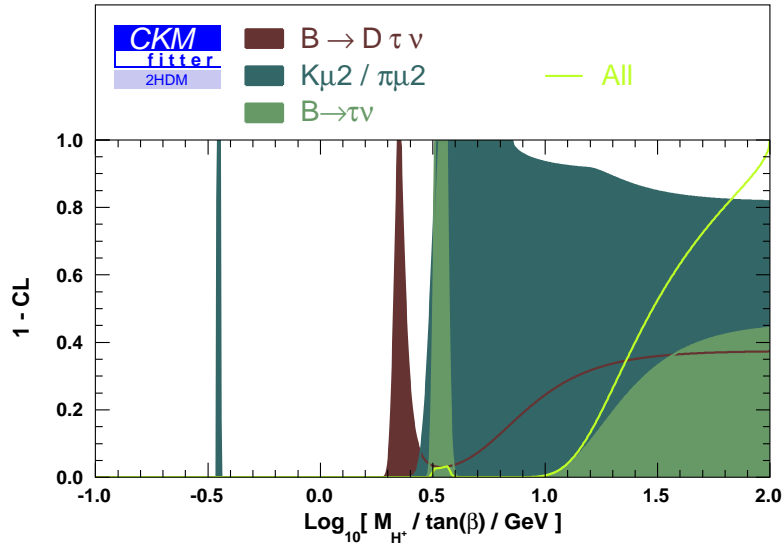,height=80mm}
\caption{\it \small Combined constraints on the ratio $m_{H^+} / \tan \beta$ derived from leptonic and semileptonic decays of $K$, $D$ and $B$ mesons, in the large $\tan \beta$
limit.
\label{fig-lepto-1D}}
\end{center}
\end{figure}

The limits derived from the semileptonic decays $B\rightarrow D \tau \nu$ and $K\rightarrow \pi \ell \nu$ are similar in shape to those derived from the purely leptonic decay
$B\rightarrow \tau \nu$, although less constraining than the latter. Hence there are not further displayed. Note that despite ${\cal  B}(B\rightarrow D \tau \nu)$ suffers from larger
theoretical and experimental uncertainties, because of the heavier mesons involved in the decay, it is 2 times more constraining on 2HDM(II) parameters than $K\rightarrow \pi \ell
\nu$.    
\par
It is very illustrative to compare all constraints derived from leptonic and semileptonic decays in the large $\tan \beta$ limit ($\tan \beta \simgt 30$). In this
limit, the charged Higgs contributions depend on a single coupling parameter, namely $\tan \beta / m_{H^+}$. Figure~\ref{fig-lepto-1D} shows the limits on $m_{H^+} / \tan \beta$ derived
from the most challenging observables, i.e., $B\rightarrow \tau \nu$, $K_{\mu 2}/\pi_{\mu 2}$ and $B\rightarrow D \tau \nu$. We also show the combined
limit from all leptonic and semileptonic decay observables considered in this study. Both $K_{\mu 2}/\pi_{\mu 2}$ and $B\rightarrow D \tau \nu$ turn out to
exclude the fine-tuned solution due to $B\rightarrow \tau \nu$, at more than $95 \%$~CL. It results in the combined limit $m_{H^+} / \tan \beta \geq
13.1$~GeV at $95 \%$~CL, from all semileptonic and leptonic decays in the large $\tan\beta$ limit.

\subsection{Loop processes}

\begin{figure}[htbp]
\begin{center}
\epsfig{file=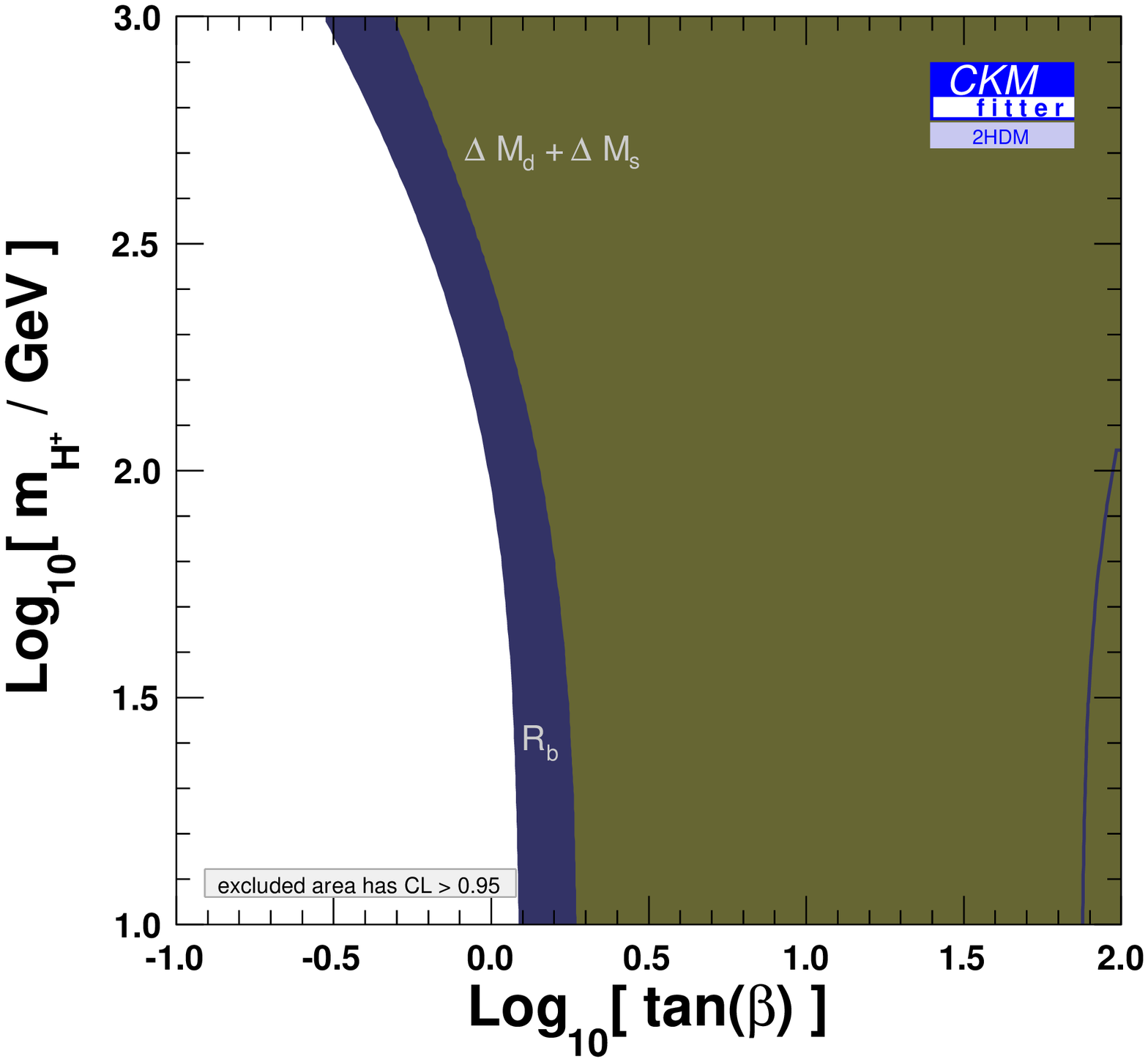,height=80mm,width=80mm}
\epsfig{file=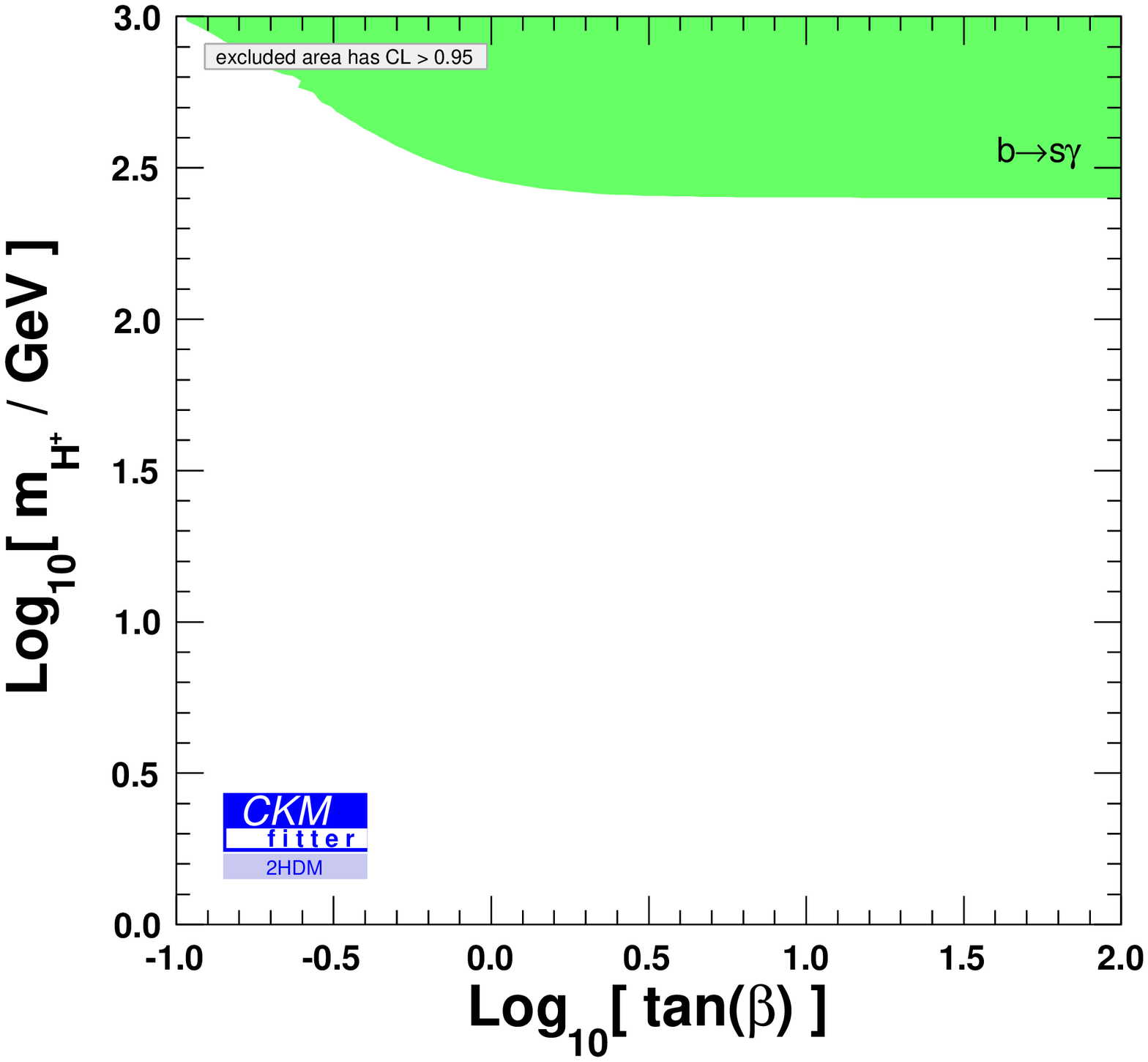,height=80mm,width=80mm}
\caption{\it \small Superimposed constraints on 2HDM parameter space ($m_{H^+}$, $\tan \beta$) from $B \bar{B}$ mixing and $Z \to b \bar{b}$ (on the left) and from 
from $b \to s \gamma$ branching ratio. \label{bsg} The colored area are confidence
regions at $95\%$~CL.\label{fig-ZMix}} 
\end{center}
\end{figure}
The constraints derived from $B\bar{B}$ mixing and $Z \to b \bar{b}$ are similar in shape as can be seen on Figure~\ref{fig-ZMix}. The constraints derived from $\Delta m_d$ and
$\Delta m_s$ are twice as stringent as those arising from $Z \to b \bar{b}$. They all exhibit a divergence with $1 /( m_{H^+} \tan\beta )$ resulting in
small values of $\tan \beta$ to be disfavoured except at very large values of $m_{H^+}$.     

The $b \to X_s \gamma$ branching ratio, where $X_s$ denotes any charmless final state with strangeness, is measured using either semi-inclusive or inclusive method. The Heavy
Flavour Averaging Group (HFAG) proposes an average of measurements~\cite{HFAGbsg}, which is used in the present work, performed by the CLEO~\cite{cleobsg}, Belle~\cite{bellebsg}
and BaBar~\cite{babarbsg} experiments (only the reference of the latest measurements is given for each collaboration). The key point of the branching ratio determination is the correction
implied by the photon energy threshold experimental cut and derived from a theoretical model of the photon spectrum shape. In that respect, HFAG advocates the use of the
extrapolation factors determined in~\cite{buchmullerbsg}. The average is:  
\begin{equation}
{\cal B}(b\rightarrow s \gamma) =(3.52 \pm 0.23 ({\rm stat}) \pm 0.09 \; ({\rm syst})) \; 10^{-4}.
\end{equation}

The $b \to s \gamma$ branching ratio is currently the constraint which dominates the global 2HDM Type II fit.
Figure~\ref{bsg} shows the exclusion region at 95\% CL in the parameter space ($m_{H^+}, \tan \beta$). This result is in fair agreement with other
determinations~\cite{bsgammaHchlimit}. Let us mention that two critical parameters in the determination of this limit are the charm quark mass $\overline{m}_c(m_c)$ and the
semileptonic phase space factor C. As an illustration, if the central value and uncertainties of $m_c$ are varied from the one quoted in Table~\ref{tab-inputs-par} to the one used in~ref.~\cite{bsgammaNNLO} ($\overline{m}_c(m_c) =(1.224 \pm 0.017 \pm
0.054) \; {\rm GeV}$), the limit on $m_{H^+}$ is increased by 10\%. Similarly for the parameter C, if to use the central value and errors from~ref.~\cite{bsgammaNNLO} ($C = 0.58
\pm 0 \pm 0.016$) the limit on $m_{H^+}$ is decreased by 9\%,
despite the error on C is two times smaller than the one we used in our analysis.

\section{Combined Analysis}

\subsection{Goodness of fit}

In order to test the relevance of the 2HDM Type II all observables are compared to their theoretical predictions within a combined global $\chi^2$ test. The $\chi^2$ is
minimized over all theory parameters, and theoretical uncertainties are treated in the \rfit\ scheme as for the global CKM fit~\cite{ThePapII}. There are $76$ observables (the determination of the $\alpha$ angle corresponds to 41 observables alone) for 
$56$ parameters, yielding $\chi^2_{min,Obs}({\rm 2HDM}) = 22.54$. From a Monte-Carlo toy-experiment study, the corresponding $p$-value is found to be~$p({\rm 2HDM})=(64.8
\pm 0.8) \%$, assuming for the true parameter values those found during the initial $\chi^2$ minimisation (plug-in $p$-value scheme~\cite{ThePapII}). It is worth comparing the
observed $\chi^2_{min}$ value in the 2HDM scheme to the one obtained for the same observables but with their SM predictions. For the latter, the $\chi^2_{min, Obs}$ is
$\chi^2_{min, Obs}({\rm SM}) = 22.56$ and a corresponding $p$-value $p({\rm SM})=(69.1 \pm 0.8) \%$ is found. Let us stress the $\chi^2_{min}$ value can only decrease when we
move from the SM to 2HDM predictions: the observables are the same but the 2HDM Type II has additional parameters which can reproduce the SM predictions in the particular limit
$m_{H^+} \rightarrow \infty$. Therefore, by comparing the observations to the predictions alone, one cannot reject or nullify the 2HDM Type II while keeping the SM. Nevertheless,
the almost equal value of $\chi^2_{min}$ for both models leads us to the qualitative conclusion that the 2HDM Type II does not perform significantly better than the SM. 
\par 
We can measure more quantitatively how the agreement with the data improves once one moves from the SM to the 2HDM Type II. For this purpose, we introduce a new test statistics,
$\Delta\chi^2_{min}$, defined as: 

\begin{align} \label{eq-DeltaChi2}
\Delta\chi^2_{min} = \chi^2_{min}( {\rm SM} ) - \chi^2_{min}( {\rm 2HDM} ).
\end{align}
       
\begin{figure}
\begin{center}
\epsfig{file=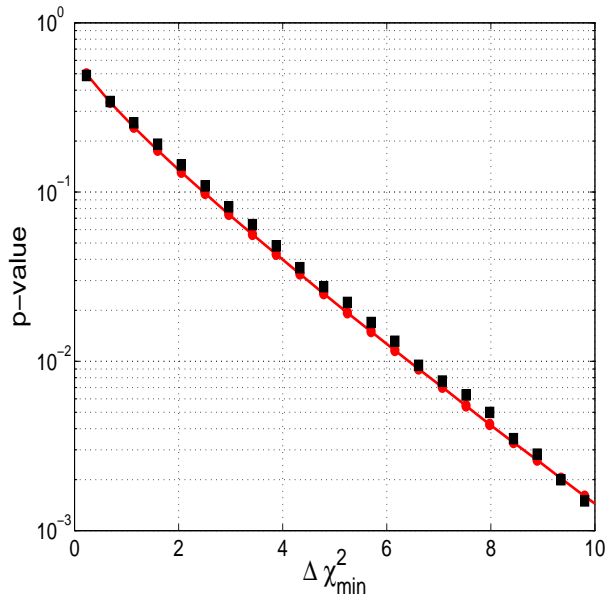,height=80mm,width=80mm}
\caption{\it \small $\Delta\chi^2_{min}$ cumulative distribution within the SM. The black squares stand for the distribution obtained from Monte-Carlo assuming the SM as the
truth. For comparison, the red dots show the distribution obtained for a 1 degree of freedom $\chi^2$ distribution with identical binning. \label{fig-DeltaChi2}} 
\end{center}
\end{figure}

High values of $\Delta\chi^2_{min}$ would indicate a deviation of the observation from the SM prediction that could be accommodated by 2HDM Type II. The cumulative distribution of
this test is derived from a Monte-Carlo toy-experiment study where the SM parameter values have been fixed to the values found in the global minimisation (this is the truth
hypothesis to nullify). The result of this toy analysis is shown in Figure \ref{fig-DeltaChi2}. A departure of $\Delta\chi^2_{min}$ at three standard deviations (a $p$-value of
$0.27\%$) would correspond to $\Delta\chi^2_{min} \geq 9.2$. The observed value $\Delta\chi^2_{min} = 0.02$ amounts to a $p$-value of $\simeq 100 \%$. Therefore, the toy analysis does not
give any reason to reject the SM in favour of the 2HDM Type II. Conversely, the 2HDM Type II hypothesis is not invalidated and the next section is dedicated to the derivation of
the exclusion limits on the model parameters ($m_{H^+},\tan \beta$).  

Eventually, we point out that the test distribution was found to be well approximated by a $\chi^2$ distribution with one degree of freedom while one would have naively expected
two degrees of freedom brought by the two additional free parameters $m_{H^+}$ and $\tan \beta$. This could be understood from the fact that the fit is dominated by the $b
\rightarrow s \gamma$ constraint, almost one-dimensional and depending only on $m_{H^+}$.

\subsection{Combined limits on 2HDM Type II parameters}

\begin{figure}
\begin{center}
\epsfig{file=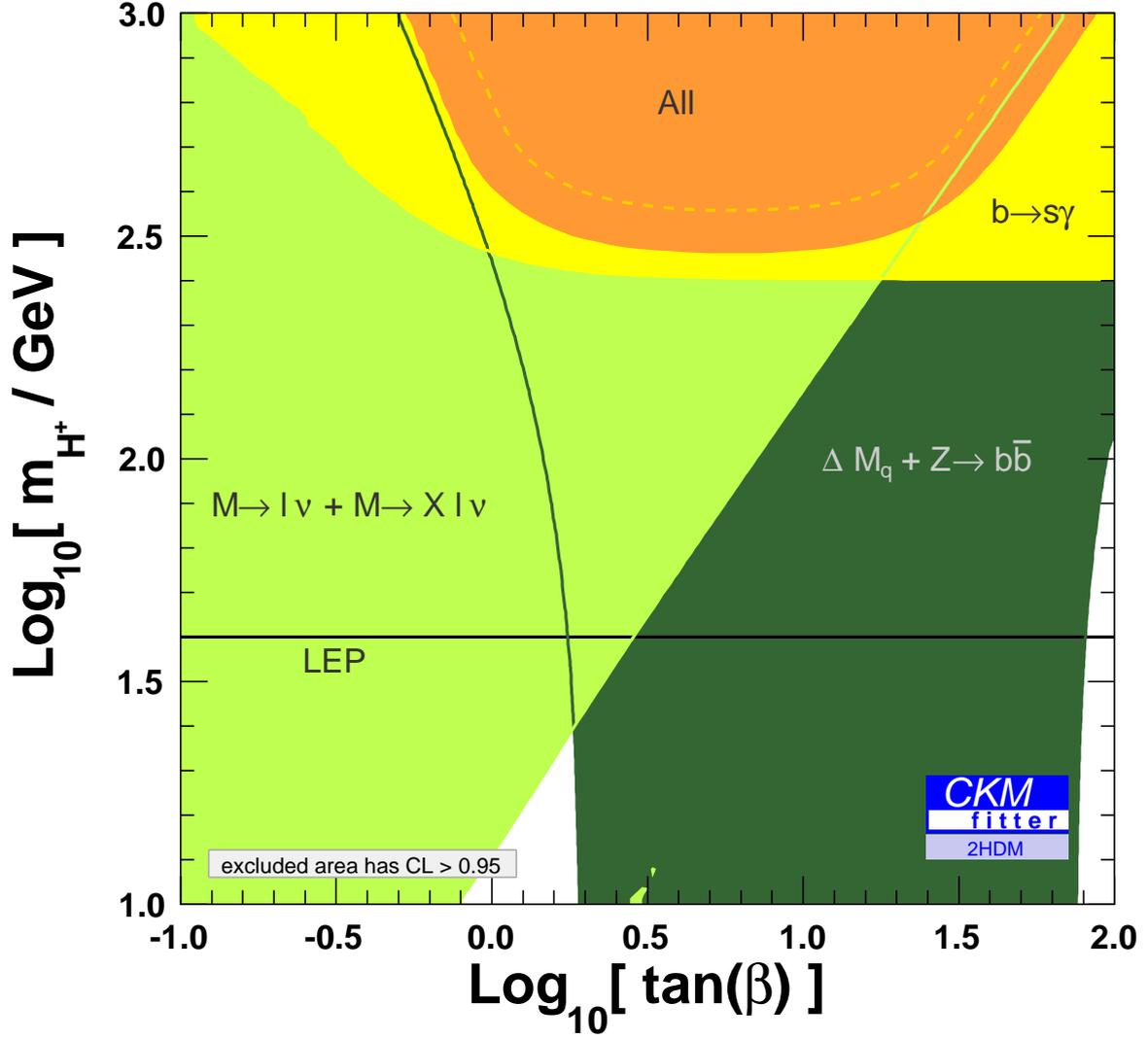,height=160mm,width=160mm}
\caption{\it \small Global constraints on 2HDM parameters $m_{H^+}$ and $\tan \beta$ from all analysed observables. Each color corresponds to a different set of observables, as
quoted in the Figure. The complementary area of the colored one is excluded at $95\%$~CL. The horizontal black line indicates the $95\%$~CL limit from direct searches at
LEP~\cite{LEPChargedHiggs}. The dotted line within the orange combined area delimits the corresponding $1 \sigma$ confidence area. For the combination of leptonic and semileptonic constraints (light
green area) we assumed that the p-value is best approximated by a 1 d.o.f. $\chi^2$ distribution since 2HDM contributions essentially depend on the ratio $m_{H^+}/\tan(\beta)$. 
\label{fig-combined}}
\end{center}
\end{figure}

 Figure~\ref{fig-combined} shows the combined $95\%$~CL confidence area in the plane ($m_{H^+}$, $\tan \beta$). The minimum $\chi^2$ of $\chi^2_{min} \simeq 22.5$ is obtained for $m_{H^+} \simeq
 4$~TeV. At high $H^+$ masses and irrespective of the value of $\tan \beta$ (decoupling limit), the charged Higgs contribution becomes negligible for all the processes we are
 considering in this analysis, so that the SM predictions are recovered.

At large $\tan \beta$ ($\simgt 30$), leptonic (mainly $B \rightarrow \tau \nu$) and semileptonic constraints compete with $b\rightarrow s \gamma$ and sharpen its exclusion limit.
At small $\tan \beta$ ($\simlt 1$), the most stringent constraint arises from the $B\bar{B}$ mixing and to a second extent from $Z \rightarrow b \bar{b}$. These results can be
compared with those obtained from the Gfitter 
group~\cite{Gfitter}, which performed a global fit to electroweak precision data both in the Standard Model and in the 2HDM Type II. In the latter case, the observables involved
were $B\to \ell \nu$, $B\to D\tau\nu$ and a ratio involving kaon decays (namely $K_{\mu 3}$, $K_{\mu 2}$ and $\pi_{\mu 2}$), $b\to s\gamma$, $Z\to b\bar{b}$ (with a more dedicated
treatment of the latter observable than in our work). The values of CKM matrix elements involved in flavour observables were taken as external inputs, whereas we determine them
from the fit described in sec.~\ref{sec:inputs}. We notice rather similar exclusion areas for the various individual constraints (e.g., the existence of fine-tuned solutions for
leptonic decays), apart from a slightly different shape in the case of $Z\to b\bar{b}$. As in our case, $b\to s\gamma$ favours high values of the charged Higgs mass, irrespective
of the value of $\tan\beta$. 

\begin{figure}
\begin{center}
\epsfig{file=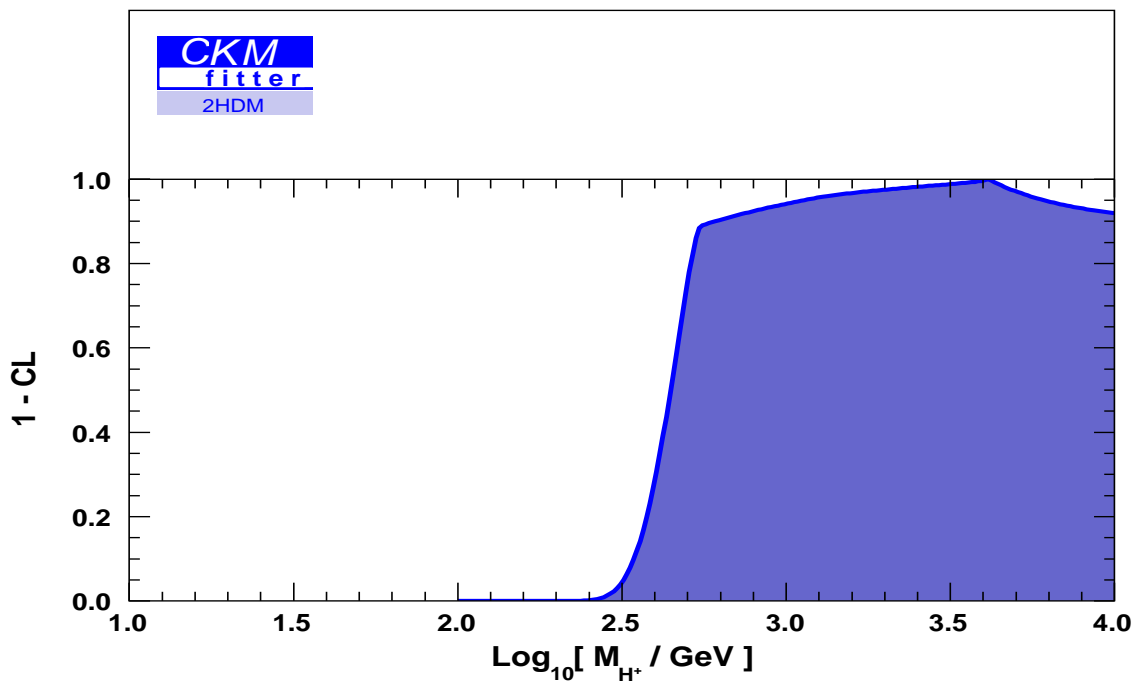,height=100mm,width=160mm}
\caption{\it \small One-dimension constraint on $m_{H^+}$.\label{1Dlnm}}
\end{center}
\end{figure}

Figure~\ref{1Dlnm} shows the one-dimension constraint found in the global analysis. A lower limit of the charged Higgs mass can be inferred: 

$$m_{H^+} > 316 \; {\rm GeV} \; {\rm at} \;  95\% \;{\rm CL} \qquad [{\rm this\ work}]$$
while no significant constraint is obtained for $\tan\beta$.

It is interesting to compare these results with the bound obtained from direct searches at LEP for any value of $\tan\beta$~\cite{LEPChargedHiggs} (specific studies are also
reported in refs.~\cite{lepcdf}): 

$$m_{H^+} > 78.6 \; {\rm GeV} \; {\rm at} \;  95\% \;{\rm CL} \qquad [{\rm direct}]$$

\section{Conclusion}

In the last decade, the $B$ factories have performed a set of remarkable measurements in the quark sector which yielded an impressive overall agreement between the SM and the
data. Tight constraints on the theories beyond the SM can be inferred: flavour-changing neutral currents are small as predicted in the SM and the KM mechanism has proven to
describe the observed CP-violating phenomena in flavour physics with a good accuracy. Among the theoretical extensions of the SM, the addition of second scalar doublet (2HDM Type
II) is particularly appealing, so that the flavour structure of the SM is preserved.
\par
In this article, we have discussed tests of the 2HDM Type II in the light of recent flavour physics data. In particular, the measurements of purely leptonic decays of $B$ and $D$ mesons (which departed more or less significantly from their SM predictions) have been combined to obtain a comprehensive combined constraint. They have been analyzed together
with complementary constraints such as semileptonic decays, $b \to s \gamma$ or $R_b$ measurements.  
\par             
The outcome of this combined analysis is that the 2HDM Type II is not favoured by low-energy data due to the interplay between ${\cal B}[\bar B \rightarrow \tau^+ \nu]$ and
$b\rightarrow s \gamma$ measurements, or  at least, that it does not perform better than the SM. If we assume that the 2HDM Type II is realized in Nature, constraints on its
parameters can be derived from this global analysis. In that respect, a limit on the charged Higgs mass $m_{H^+} > 316 \; {\rm GeV}$ at $95\%$~CL is obtained irrespective of $\tan \beta$.
\par 

This analysis considered tree and loop-induced $\Delta F = 1$ processes,
as well as  $\Delta F = 2$ mixing processes for neutral $B$ mesons.  Loop-induced rare decays such as $B_s \to
\mu^+ \mu^-$ might receive contributions from an extended Higgs sector, and would provide further constraints on this model. A natural extension of this work would include such
observables in order to perform a more comprehensive test of the Two Higgs Doublet Model of Type II. 

\section*{Acknowledgments}

We thank our colleagues of the CKMfitter group for many stimulating discussions and for a critical reading of the manuscript. We also thank J. Orloff and N. Mahmoudi for very useful comments on two Higgs doublet models. We eventually thank H. Fl\" acher and A. H\" ocker (from the Gfitter group) for valuable discussions on the constraints and P. Slavich and G. Degrassi for their useful comments on the $R_b$ observable description. This work was
supported in part by the ANR contract ANR-06-JCJC-0056 and by EU Contract No. MRTN-CT-2006-035482, \lq\lq FLAVIAnet''. 

\appendix

\section{Parametrization for $K^0\rightarrow \pi \ell \nu$}

The phase space integrals $\hat{I}_K^{\ell}$ in eq.~(\ref{eq-KpiGlobal}) is split into two integrals as $\hat{I}_{K}^{\ell}=\hat{I}_{K,+}^{\ell}+\hat{I}_{K,0}^{\ell}$:
\begin{align}
\hat{I}_{K,+}^{\ell} & = \int_{r_{\ell}}^{(1-\sqrt{r_\pi})^2}\lambda^3[u](1+\frac{r_{\ell}}{2u})(1-\frac{r_{\ell}}{u})^2 \left| \hat{f}_{+}[u] \right|^2  \label{eq-KpiIntegral-1}, \\
\hat{I}_{K,0}^{\ell} & = \frac{3}{2}(1-r_{\pi}^2)r_{\ell}\int_{r_{\ell}}^{(1-\sqrt{r_\pi})^2}\frac{\lambda[u]}{u}(1-\frac{r_{\ell}}{u})^2 \left| \hat{f}_{0}[u] \right|^2, \label{eq-KpiIntegral-2}
\end{align}
involving the rescaled transfer momentum $u=t/m_{K^0}^2$, the rescaled masses $r_{\pi}=(m_\pi/m_{K^0})^2$, $r_{\ell}=(m_{\ell}/m_{K^0})^2$ and the phase space term
$\lambda[u]=\sqrt{(u - (1+r_\pi)^2)-4r_\pi^2}$.  

A representation of the two $K\pi$ form factors is needed to compute these integrals.
The vector form factor $\hat{f}_{+}$ is not sensitive to charged Higgs
contributions. We use a pole parametrization with pole mass $M_V$ tuned on experimental data. For the scalar form factor, which is sensitive to $H^+$ contributions we follow the
dispersive parametrization given in~ref.~\cite{scalarFF}. This is based on a dispersion relation proposed with two subtraction points where the scalar form factor is well known:
the normalization at 0 (determined from lattice QCD) and the Callan-Treiman point (at $t=m_K^2-m_\pi^2$) where it is predicted to be related to $f_K/f_\pi$ up to very small
corrections. It requires the knowledge of $\pi K$ scattering phase shifts at low energies, obtained through the analysis of Roy-Steiner equations~(ref.~\cite{piK}). 
\begin{equation} \label{eq-KpiFormFactors}
\hat{f}_{+}[u]  = \frac{r_V}{r_V-u}, \qquad
\hat{f}_{0}[u]  = \exp\left(\frac{u}{1-r_\pi}\left(z-G\left[\frac{u}{(1-\sqrt{r_\pi})^2}\right]\right)\right),
\end{equation}
with the rescaled mass of the vector resonance $r_V=(M_V/m_{K^0})^2$, and the subtraction constant of the dispersion relation 
\begin{equation}
z=\hat{f}_0=f_K/f_\pi/f_+(0)+\Delta_{CT}- \frac{m_s \tan^2\beta-m_u}{m_s-m_u}\left(\frac{m_{K^0}^2-m_{\pi^+}^2}{m_{H^+}^2}\right)
\end{equation}
where charged Higgs contributions are included. 

The functional dependence of $G$ is:
\begin{align} \label{eq-KpiScalar}
G[x]\cdot10^2 & = (2.09+0.26\epsilon_G) x + ( 3.98 + 0.56\epsilon_G )(1- x) + (0.45+0.01\epsilon_G) x(1-x),
\end{align}

where we introduced theoretical uncertainties through the parameter $\epsilon_G$ lying in $[-1;1]$. Substituting the form factor expressions in eqs.~(\ref{eq-KpiIntegral-1}) and~(\ref{eq-KpiIntegral-2}), the ratios of phase space integrals are computed by numerical integration for various values of the parameters $z \in [-0.4;0.8]$, $\epsilon_G \in [-1;1]$.
The relative variations of the integrals ratio propagate linearly with $\epsilon_G$ and $\Delta M_V$ over the investigated range. The local derivative with respect to the latter
parameters only depend on $z$ and their variations are smooth. These observations led us to the parameterization given in eq.~(\ref{eq-KpiParam}), in perfect agreement with our
numerical studies.    

\section{Parametrization for $b\rightarrow s \gamma$}

\subsection{Analytic expressions}

At the order of leading logarithms, the relevant Wilson coefficient for the effective
SM Hamiltonian, $C_{7,SM}^{eff,(0)}$, of eq.~(\ref{eq-bsgammaPara}) writes as:
\begin{align} \label{eq-bsgammaLL}
C_{7,SM}^{eff,(0)} = \eta^{\frac{16}{23}}C_{7,SM}^{(0)}[\mu_0]+\frac{8}{3}(\eta^{\frac{14}{23}}-\eta^{\frac{16}{23}})C_{8,SM}^{(0)}[\mu_0]+\sum_{i=1}^{8}{h_i\eta^{a_i}},
\end{align}	     
where $\eta=\alpha_s[\mu_0]/\alpha_s[\mu_b]$ is the ratio of the strong coupling constants at the scales $\mu_0$ and $\mu_b$. We followed ref.~\cite{SusyBSG} by setting
$\mu_0=2 m_W$ and $\mu_b=1.5$~GeV as the mass scales for the renormalization scheme. The Wilson coefficients $C_{7,SM}^{(0)}$ and $C_{8,SM}^{(0)}$, at scale $\mu_t$, write:
\begin{align} \label{eq-Wilson}
C_{7,SM}^{(0)}[\mu_0] & = -\frac{x_{tW}}{2}( 2 F_1[x_{tW}] + 3 F_2[x_{tW}] ), \\
C_{8,SM}^{(0)}[\mu_0] & = -\frac{3 x_{tW}}{2} F_1[x_{tW}], 
\end{align}	   
with $x_{tW}=(m_t[\mu_0]/m_W)^2$ and the functions $F_i$:
\begin{alignat}{2} \label{eq-F12}
F_1[x] & = \frac{x^3-6x^2+3x+2+6x \ln(x)}{12(x-1)^4}, \\
F_2[x] & = \frac{2x^3+3x^2-6x+1-6x^2 \ln(x)}{12(x-1)^4}. 
\end{alignat} 
The running of $\alpha_s$ is computed at the second order with $n_f=5$ active flavours. One has:
\begin{align} \label{eq-alphaS}
\alpha_s^{(5)}[ \mu, \Lambda_{QCD} ] = \frac{ 12 \pi }{ 23 l_{ \mu \Lambda } } \left( 1 - \frac{ 348 }{ 529 } \frac{ \ln( l_{ \mu \Lambda } ) }{ l_{ \mu \Lambda } } \right),  
\end{align}	   
with $l_{ \mu \Lambda } = 2 \ln( \mu/\Lambda_{QCD} )$ and the matching condition $\alpha_s^{(5)}[ m_Z, \Lambda_{QCD} ] = \alpha_s(m_Z)$. Moreover, the running quark mass,
$m_t[\mu_0]$, at scale $\mu_0$ is computed at $3^{rd}$ order in $\alpha_s$ from the pole quark mass, following \cite{RunningQuarkMass}. For $n_f=5$ active flavours, it goes as: 
\begin{alignat}{2} \label{eq-Running}
m_t[\mu] & = m_t^{\rm pole} ( 1 - \left(  \frac{4}{3} + l_{\mu M} \right) a_{\mu} - \left( 9.125 + \frac{419}{72} l_{\mu M} + \frac{2}{9} l_{\mu M}^2 \right) a_{\mu}^2 \\
         & - \left( 0.3125 l_{\mu M}^3 + 4.5937 l_{\mu M}^2 + 25.3188 l_{\mu M} + 81.825 \right) a_{\mu}^3 )
\end{alignat}
with $l_{\mu M} = 2 \ln( \mu/m_t^{\rm pole} )$ and $a_{\mu} = \alpha_s^{(5)}[ \mu, \Lambda_{QCD} ] / \pi$.
Finally, the third term in eq.~(\ref{eq-bsgammaLL}) further accounts for the remaining Wilson coefficients $C_{i=[1;6]}$. We collected these relevant numbers in 
Table~\ref{tab-bsgammaLL}.
\begin{table}
\begin{center}
\begin{tabular}{c|*{9}{c}}
  \hline
  \hline
     $i$ & $1$ & $2$ & $3$ & $4$ & $5$ & $6$ & $7$ & $8$ \\
  \hline
  \hline
     $h_i$  & $\frac{626126}{272277}$ & $-\frac{56281}{51730}$ & $-\frac{3}{7}$ & $-\frac{1}{14}$  & $-0.6494$ & $-0.0380$  & $-0.0185$ & $-0.0057$ \\
     $a_i$  & $\frac{14}{23}$         & $\frac{16}{23}$        & $\frac{6}{23}$ & $-\frac{12}{23}$ & $0.4086$  & $-0.42230$ & $-0.8994$ & $0.1456$  \\
  \hline
  \hline
\end{tabular}
\caption{\small \it Magic numbers required to evaluate the LL perturbative contributions to $b\rightarrow s$, taken from~ref.~\cite{bsgammaQuark}.\label{tab-bsgammaLL}}
\end{center}
\end{table}

Charged Higgs contributions play in at leading logarithm order in the Wilson coefficients $C_{7}^{(0)}$ and $C_{8}^{(0)}$ by substituting the $W^\pm$ boson with the charged Higgs
$H^\pm$. They result in  
the substitution in eq.~(\ref{eq-bsgammaLL}): $C_{i,SM}^{(0)} \rightarrow C_{i,SM}^{(0)} + \Delta C_{i,H^+}^{(0)}$, with:
\begin{alignat}{2} \label{eq-WilsonHch}
\Delta C_{7,H^+}^{(0)}[\mu_0] & = -\frac{x_{tH}}{2} \left( \frac{1}{\tan^2\beta}( \frac{2}{3} F_1[x_{tH}] + F_2[x_{tH}] ) + \frac{2}{3} F_3[x_{tH}] + F_4[x_{tH}] \right), \\
\Delta C_{8,H^+}^{(0)}[\mu_0] & = -\frac{x_{tH}}{2} \left( \frac{F_1[x_{tH}]}{\tan^2\beta} + F_3[x_{tH}] \right),
\end{alignat}		 
where $x_{tH}=(m_t[\mu_0]/m_{H^+})^2$ and $F_{3,4}$ is defined as:
\begin{equation} \label{eq-F34}
F_3[x]  =  \frac{x^2-4x+3+2\ln(x)}{2(x-1)^3}, \qquad
F_4[x]  = \frac{x^2-1-2x \ln(x)}{2(x-1)^3}. 
\end{equation}		 
Following, the charged higgs contributions, $\Delta C_{7,H^+}^{eff,(0)}$, to the effective Wilson coefficient fall from eq.~(\ref{eq-bsgammaLL}), as:
\begin{align}
\Delta C_{7,H^+}^{eff,(0)} = \eta^{\frac{16}{23}}C_{7,H^+}^{(0)}[\mu_0]+\frac{8}{3}(\eta^{\frac{14}{23}}-\eta^{\frac{16}{23}})C_{8,H^+}^{(0)}[\mu_0].
\end{align}	     

The higher-order perturbative terms contain charm quark contributions and electroweak effects in addition. Hence, they also 
depend on the running quark mass $\overline{m}_c(m_c)$ and on the photon energy experimental cut-off, $E_0\simeq$ 1.6 GeV. They are known exactly at NLO and 
parameterized at NNLO \cite{bsgammaNNLO}. A full expression for the NLO term is very lengthy, as can be inferred starting from the final results 
given in ref.~\cite{bsgammaNNLO}. The non perturbative corrections seem to be only partly known \cite{bsgammaQuark}-\cite{bsgammaNNLO}. They are small as 
compared to other terms (a $~10\%$ effect) and uncertainties mainly depend on $m_c$ \cite{bsgammaQuark}.     

\subsection{Parametric model}

Perturbative ($P$) and non-perturbative ($N$) contributions are modeled according to eq.~(\ref{eq-bsgammaPara}) with two parametric functions $A$ and $B$, depending on a subset of
input parameters: $m_t^{pole}$, $\overline{m}_c(m_c)$ and $\alpha_s(m_Z)$. We neglect the dependence on the photon cut-off energy, which is not quoted in ref.~\cite{bsgammaNNLO}.
Following our assumptions, the values of $A$ and $B$ were computed for different sets of parameters by fitting the $\bar{B}\rightarrow X_s\gamma$ branching ratio given by SusyBSG
as a function of the charged Higgs mass, $m_{H^+}$. We found perfect agreement between SusyBSG results and our model, which justifies {\it a posteriori} the expression given in
eq.~(\ref{eq-bsgammaPara}) and our parametric assumptions. 

We further studied the variations of $A$ and $B$ as a function of the inputs $m_t^{{\rm pole}}$, $\overline{m}_c(m_c)$ and $\alpha_s(m_Z)$. We cross-checked that the variations of
$A$ and $B$ with respect to the others  parameters, quoted in Table~4 of ref.~\cite{bsgammaNNLO}, are negligible. Those with respect to $m_t^{{\rm pole}}$, $\overline{m}_c(m_c)$ and
$\alpha_s(m_Z)$ are linear for small changes. We notice that linearity holds for deviations of the input parameters as strong as $3\sigma$ from their central value. Therefore, it
seems very appropriate to approximate the functions $A$ and $B$ by their leading-order Taylor expansions around the central input parameter values: 
\begin{alignat}{2} \label{paramAB}
A & = A_0(1+a_c\Delta \overline{m}_c(m_c) + a_t\Delta m_t^{{\rm pole}} + a_s\Delta \alpha_s(m_Z)), \\
B & = B_0(1+b_c\Delta \overline{m}_c(m_c) + b_t\Delta m_t^{{\rm pole}} + b_s\Delta \alpha_s(m_Z)), 
\end{alignat}		 
where $\Delta p$ denotes the variation of the input parameter $p$ around its central value. The coefficients values were extracted from a polynomial fit, summarized in
Table~\ref{tab-param}. 

\begin{table}
\begin{center}
\begin{tabular}{c|cccc}
  \hline
  \hline
        & $X_0$ & $x_c$ & $x_t$ & $x_s$ \\
  \hline
  \hline
     A  & $3.155 \cdot 10^{-2}$ & $2.80$                & $-1.06 \cdot 10^{-4}$ & $36.2$ \\
     B  & $7.564 \cdot 10^{-1}$ & $-2.43 \cdot 10^{-1}$ & $-7.68 \cdot 10^{-4}$ & $-4.62$ \\
  \hline
  \hline
\end{tabular}
\caption{\small \it Coefficients of the Taylor expansion of $A$ and $B$ functions. The label X stands for A or B. \label{tab-param}}
\end{center}
\end{table}

\section{Parametrization for $Z\rightarrow b\bar{b}$}

\subsection{SM prediction}

Following~\cite{EWFit}, we performed a global Electroweak fit to the Z pole observables in the High-$Q^2$ scheme, assuming purely Gaussian
errors and taking their correlations into account. There are five fit parameters: the masses of the $Z$ boson, the neutral Higgs boson and the top quark
as well as the strong coupling constant, $\alpha_s[m_Z]$, and the hadronic vacuum polarization, $\Delta\alpha_{had}^{(5)}[m_Z]$, considered at the $Z$ pole.
The values of the $Z$ pole observables and the hadronic vacuum polarization are taken from~\cite{EWFit}(table 8.4) but excluding the measurements of $R_b$, $A_{FB}^{0,b}$ and $A_b$. 

The top quark mass is essentially constrained by the direct measurement given in Table~\ref{tab-inputs-par} while the $W$ boson mass
and width are taken from~\cite{PDG}. With these inputs we get: $R_b=0.21580(4)$ and $\log_{10}(m_{H^0}/{\rm GeV})=1.74123 \pm 0.17982$~\footnote{It has been checked that the global EW fit results~\cite{EWFit} were accurately reproduced when taking all inputs into account.}.

The shape of the Electroweak fit $\Delta \chi^2$ around its global minimum follows a 
Gaussian distribution. Therefore, it is modelled by a
single observable $\log_{10}(m_{H^0}/{\rm GeV})$, linearly correlated to $m_t^{pole}$ and
$\Delta\alpha_{had}^{(5)}$, while the two remaining parameters, $m_Z$ and
$\alpha_s[m_Z]$ are considered constant. The linear correlation factors
$r[\log_{10}(m_{H^0}),\Delta\alpha_{had}^{(5)}]=0.261$ and $r[\log_{10}(m_{H^0}),m_t^{pole}]=-0.494$ are derived from the shape of the
multidimensional $\Delta\chi^2$ in the neighbourhood of its global minimum, as:
\begin{align} \label{eq-correlation}
  C[x,y] = \frac{\partial y^{*}}{\partial x} \frac{\sigma_x}{\sigma_y}, 
\end{align}    
where $y^{*}[x]$ is the parameter value that minimises the $\Delta \chi^2$ for the fixed value of $x=\log_{10}(m_{H^0}/{\rm GeV})$
and where the derivative is taken at the best guess value $x^{*}$, the one yielding the global minimum.

The left and right couplings, $\bar{g}_L^b$ and $\bar{g}_R^b$, are expressed as a function of $\sin^2\theta^b_{\rm eff}$ and the Veltman's parameter $\rho_b$~:
\begin{align} \label{eq-Veltman}
\bar{g}_L^b = \sqrt{\rho_b}(T_3^b-Q_b \sin^2\theta^b_{\rm eff}), \\
\bar{g}_R^b = -\sqrt{\rho_b} Q_b \sin^2(\theta^b_{\rm eff}),
\end{align}
whith $T_3^b=1/2$ and $Q_b=-1/3$ the third component of the weak isospin and the electric charge of the $b$ quark. The SM parameters $K_b$, $\sin^2(\theta^b_{\rm eff})$ and $\rho_b$ are thus
parameterized by polynomials of the 3 relevant input parameters of the Electroweak fit $m_t^{pole}$, $\log_{10}(m_{H^0})$ and $\Delta\alpha_{\rm had}^{(5)}$, in
order to reproduce ZFITTER results~\cite{ZFITTER}. Variations with $m_t^{pole}$ and $\Delta\alpha_{\rm had}^{(5)}$ are linear up to $5\sigma$ from their expectation. 
Variations with $\log_{10}(m_{H^0})$ could however only be reproduced by a $4^{th}$ order polynomial. The parameterization is written as:
\begin{align}
\label{eq-Rb-param}
\begin{pmatrix}
K_b \cr
\sin^2 \theta^b_{\rm eff} \cr
\rho_b \cr
\end{pmatrix} = 
A_{Z\rightarrow b\bar{b}}
\begin{pmatrix}
\Delta_t \cr
\Delta_{\alpha} \cr
\Delta_H \cr
\Delta_H^2 \cr
\Delta_H^3 \cr
\Delta_H^4 \cr
\end{pmatrix} + 
\begin{pmatrix}
    1.333120 \cr
    0.232641 \cr
    0.994306 \cr
\end{pmatrix}, 
\end{align}
where $\Delta_X$ are the variations of the parameters $m_t^{pole}$, $\Delta\alpha_{\rm had}^{(5)}$ and $\log_{10}(m_{H^0})$ (denoted $X=t,\alpha,H$), around their central values (respectively $172.4$~GeV, $0.02758$ and $1.74123$). $A_{Z\rightarrow b\bar{b}}$ is a $3\times 6$ matrix whose coefficients are given in the transposed matrix of eq.~\ref{A-param}. With these numerical values one reproduces the predictions for $K_b = 1.33312$, $\sin^2\theta^b_{\rm eff} = 0.232641$ and $\rho_b = 0.994306$ to a relative accuracy of $10^{-6}$, for values of the input parameters within a $3 \sigma$ range from their central values.

\begin{align}
\label{A-param}
A^t_{Z\rightarrow b {\bar b}} = 
\begin{pmatrix}[r]
 1.97262 \cdot 10^{-4}  & -6.07684 \cdot 10^{-6}  &  -7.34121 \cdot 10^{-5} \cr 
   -0.685225            & 0.348494              & 1.73585 \cdot 10^{-3}    \cr
 -1.32964 \cdot 10^{-3}  &  8.52074 \cdot 10^{-4}  & 6.36640 \cdot 10^{-4}   \cr
 -3.91140 \cdot 10^{-3} &  1.57997 \cdot 10^{-4}  & -2.59313 \cdot 10^{-3}  \cr
 9.57438 \cdot 10^{-4}   & 0                      &  3.99581 \cdot 10^{-4}   \cr
 2.62977 \cdot 10^{-4}   &  0                     & 2.56867 \cdot 10^{-4}    \cr
\end{pmatrix}. 
\end{align}


\subsection{Charged Higgs contribution}

The charged Higgs contributions can be implemented through a redefinition of the coupling constants, as given in Eq.~(\ref{eq-gbLR}), where the function $F_z$, at 2 loops is: 
\begin{align} \label{eq-Fz}
F_z[x] = f_1[x] + \frac{\alpha_s(\mu_0)}{3\pi}f_2[x]. 
\end{align}
with $x=m^2_t(\mu_0)/m^2_{H^+}$. Computations are performed at the scale $\mu_0=2m_W$ as for $b\rightarrow s \gamma$, in Appendix~B. The two loops correction, $f_2$, was computed in ref.~\cite{Slavich}. To avoid di-logarithm computations, we parameterized
the corresponding terms in $f_2$ by rational fractions, as:
\begin{align} \label{eq-f1f2}
f_1[x] = & \frac{x^2 - x - x \ln(x)}{(x-1)^2}, \\
f_2[x] = & \frac{12 f_1[x] - 6 x}{x - 1}\ln \left( \frac{m_t(\mu_0)^2}{\mu_0^2} \right) + \frac{P_2[x]}{Q_2[x]} f_1[ x ], 
\end{align}				 
where $P_2$ and $Q_2$ are $4^{th}$ order polynomials whose coefficients are given in Table~\ref{table-Rb-PQ}. The latter parameterization reproduces the results of~\cite{Slavich} with an accuracy better than $0.5\%$
 for $x$ in a range from $10^{-5}$ to $10^{5}$.  

\begin{table}
\begin{center}
\begin{tabular}{c|*{5}{c}}
  \hline
  \hline
          & $x^0$ & $x^1$ & $x^2$ & $x^3$ & $x^4$ \\
  \hline
  \hline
     $P_2$  & $~0.953342$ & $-0.197937$              & $~5.50096\cdot 10^{-2}$ & $-3.22697\cdot 10^{-3}$ & $~1.30788\cdot 10^{-4}$ \\
     $Q_2$  & $-0.219441$ & $-2.72075\cdot 10^{-2}$  & $-5.26861\cdot 10^{-3}$ & $-2.74128\cdot 10^{-4}$ & $-7.53345\cdot 10^{-6}$ \\
  \hline
  \hline
\end{tabular}
\caption{\small \it Polynomial coefficients for the parameterization of the 2HDM(II) prediction of $R_b$. Coefficients are listed in increasing power of $x$.
\label{table-Rb-PQ}}
\end{center}
\end{table}

\vspace{1.cm}


\clearpage


\begin{thebibliography}{110}


\bibitem{SM_CKM}
  M.~Kobayashi and T.~Maskawa,
  Prog.\ Theor.\ Phys.\  {\bf 49} (1973) 652.

  L.~Wolfenstein,
  Phys.\ Rev.\ Lett.\  {\bf 51}, 1945 (1983).

\bibitem{revNP}
  M.~E.~Peskin,
  arXiv:hep-ph/9705479.

  R.~D.~Peccei,
  arXiv:hep-ph/9909233.

  J.~R.~Ellis,
  arXiv:hep-ph/9812235.

  Y.~Nir,
  arXiv:hep-ph/0109090.

\bibitem{2hdmref1}
  L.~F.~Abbott, P.~Sikivie and M.~B.~Wise,
  Phys.\ Rev.\  D {\bf 21} (1980) 1393.

\bibitem{2hdmref2}
  G.~C.~Branco, A.~J.~Buras and J.~M.~Gerard,
  \ Nucl.\ Phys.\  B {\bf 259} (1985) 306.

\bibitem{2hdmref3}
  V.~D.~Barger, J.~L.~Hewett and R.~J.~N.~Phillips,
  Phys.\ Rev.\  D {\bf 41} (1990) 3421.

\bibitem{2hdmref4}
  W.~S.~Hou,
  Phys.\ Rev.\  D {\bf 48} (1993) 2342.

\bibitem{2hdmref5}
  U.~Haisch,
  [arXiv:hep-ph/08052141].

\bibitem{2hdmref6}
  A.~Pich and P.~Tuzon,
  Phys.\ Rev.\  D {\bf 80} (2009) 091702
  [arXiv:0908.1554 [hep-ph]].

 M.~Jung, A.~Pich and P.~Tuzon,
  arXiv:1006.0470 [Unknown].

\bibitem{2hdmsearch}

  K.~A.~Assamagan, Y.~Coadou and A.~Deandrea,
  Eur.\ Phys.\ J.\ direct C {\bf 4}, 9 (2002)
  [arXiv:hep-ph/0203121].

 P.~Salmi, R.~Kinnunen and N.~Stepanov,
  arXiv:hep-ph/0301166.

  A.~G.~Akeroyd and M.~Aoki,
  Phys.\ Rev.\  D {\bf 72}, 035011 (2005)
  [arXiv:hep-ph/0506176].

\bibitem{ThePapII}
  J.~Charles {\it et al.}  [CKMfitter Group],
  Eur.\ Phys.\ J.\  C {\bf 41}, 1 (2005)
  [arXiv:hep-ph/0406184].

\bibitem{2hdmexp}
  A.~G.~Akeroyd and C.~H.~Chen,
  Phys.\ Rev.\  D {\bf 75}, 075004 (2007)
  [arXiv:hep-ph/0701078].

  A.~G.~Akeroyd and S.~Recksiegel,
  J.\ Phys.\ G {\bf 29}, 2311 (2003)
  [arXiv:hep-ph/0306037].

\bibitem{Gfitter}
  H.~Flacher, M.~Goebel, J.~Haller, A.~Hocker, K.~Moenig and J.~Stelzer,
  Eur.\ Phys.\ J.\  C {\bf 60}, 543 (2009)
  [arXiv:0811.0009 [hep-ph]].

\bibitem{cleo_update}  
  J.~P.~Alexander {\it et al.}  [CLEO Collaboration],
  Phys.\ Rev.\  D {\bf 79}, 052001 (2009)
  [arXiv:0901.1216 [hep-ex]].

  P.~U.~E.~Onyisi {\it et al.}  [CLEO Collaboration],
  Phys.\ Rev.\  D {\bf 79}, 052002 (2009)
  [arXiv:0901.1147 [hep-ex]].

\bibitem{cleo_summer}
  O.~Deschamps,
  arXiv:0810.3139 [hep-ph].

\bibitem{FlaviaNet}
  M.~Antonelli {\it et al.}  [FlaviaNet Working Group on Kaon Decays],
  arXiv:0801.1817 [hep-ph]. And references therein. 
  
\bibitem{PDG} 
  C.~Amsler {\it et al.}  [Particle Data Group],
  Phys.\ Lett.\  B {\bf 667}, 1 (2008).

\bibitem{CLEO}  B.~I.~Eisenstein {\it et al.}  [CLEO Collaboration],
  Phys.\ Rev.\  D {\bf 78}, 052003 (2008)
  [arXiv:0806.2112 [hep-ex]].

\bibitem{Stone}
  J.~L.~Rosner and S.~Stone,
  arXiv:0802.1043 [hep-ex].

\bibitem{BaBar}
  B.~Aubert {\it et al.}  [BaBar Collaboration],
  Phys.\ Rev.\  D {\bf 77}, 011107 (2008)
  [arXiv:0804.2422 [hep-ex]].

\bibitem{Belle}
  K.~Ikado {\it et al.}  [Belle Collaboration],
  Phys.\ Rev.\ Lett.\  {\bf 97} (2006) 251802
  [arXiv:hep-ex/0604018].

  I.~Adachi {\it et al.}  [Belle Collaboration],
  arXiv:0809.3834 [hep-ex].

\bibitem{bdtaunubelle} 
  A.~Matyja {\it et al.}  [Belle Collaboration],
  Phys.\ Rev.\ Lett.\  {\bf 99} (2007) 191807
  [arXiv:0706.4429 [hep-ex]].

\bibitem{bdtaunubabar} 
  B.~Aubert {\it et al.}  [BABAR Collaboration],
  arXiv:0902.2660 [hep-ex].

\bibitem{CDF-Dms}
  A.~Abulencia {\it et al.} [CDF Collaboration],
  Phys. Rev. Lett. {\bf 97}, 242003 (2006).

\bibitem{EWFit}
The ALEPH, DELPHI, L3 and OPAL Collaborations
  Phys.\ Rept.\  {\bf 427}, 257 (2006)
  [arXiv:hep-ex/0509008].

\bibitem{HFAGbsg} 
  E.~Barberio {\it et al.}  [Heavy Flavor Averaging Group],
  arXiv:0808.1297 [hep-ex].
  
\bibitem{TevatronEWG}
  The Tevatron Electroweak Working Group, for the CDF, D0 Collaborations,
  arXiv:0808.1089 [hep-ex].

\bibitem{lqcd_ckmfitter} 
  V.~Tisserand,
  arXiv:0905.1572 [hep-ph].

\bibitem{MesciaKamenik}
  J.~F.~Kamenik and F.~Mescia,
  Phys.\ Rev.\  D {\bf 78}, 014003 (2008)
  [arXiv:0802.3790 [hep-ph]].

\bibitem{Buchalla}  
  G.~Buchalla {\it et al.},
 Rev. Mod. Phys. {\bf 68}, 1125 (1996).

\bibitem{Lenz_etaB} 
  A.~Lenz,{/it private communication}.

\bibitem{latticebtod}
  G.~M.~de Divitiis, R.~Petronzio and N.~Tantalo,
  JHEP {\bf 0710} (2007) 062
  [arXiv:0707.0587 [hep-lat]].
  
\bibitem{bsgammaNewC}
  P.~Gambino and P.~Giordano,
  Phys.\ Lett.\ B {\bf 669}, 69 (2008)
  [arXiv:hep-ph/0805.0271].
  
\bibitem{RunningQuarkMass}
  K.~G.~Chetyrkin and M.~Steinhauser,
  Nucl.\ Phys.\ B {\bf 573}, 617 (2000)
  [arXiv:hep-ph/9911434]
  
\bibitem{Kl2pil2em}
  W.~J.~Marciano,
  Phys.\ Rev.\ Lett.\  {\bf 93} (2004) 231803
  [arXiv:hep-ph/0402299].

 V.~Cirigliano and I.~Rosell,
  JHEP {\bf 0710} (2007) 005
  [arXiv:0707.4464 [hep-ph]].

  V.~Cirigliano and I.~Rosell,
  Phys.\ Rev.\ Lett.\  {\bf 99} (2007) 231801
  [arXiv:0707.3439 [hep-ph]].

\bibitem{Dl2em}
 B.~A.~Dobrescu and A.~S.~Kronfeld,
  Phys.\ Rev.\ Lett.\  {\bf 100} (2008) 241802
  [arXiv:0803.0512 [hep-ph]].

\bibitem{2hdm2ds}
  A.~G.~Akeroyd,
  Prog.\ Theor.\ Phys.\  {\bf 111}, 295 (2004)
  [arXiv:hep-ph/0308260].

  A.~G.~Akeroyd and F.~Mahmoudi,
  JHEP {\bf 0904}, 121 (2009)
  [arXiv:0902.2393 [hep-ph]].

\bibitem{BtoDstar}
 J.~G.~Korner and G.~A.~Schuler,
  Phys.\ Lett.\  B {\bf 231} (1989) 306.

  J.~G.~Korner and G.~A.~Schuler,
  Z.\ Phys.\  C {\bf 46} (1990) 93.

  M.~Tanaka,
  Z.\ Phys.\  C {\bf 67} (1995) 321
  [arXiv:hep-ph/9411405].

  C.~H.~Chen and C.~Q.~Geng,
  JHEP {\bf 0610} (2006) 053
  [arXiv:hep-ph/0608166].


\bibitem{NiersteBDtaunu}
  U.~Nierste, S.~Trine and S.~Westhoff,
  Phys.\ Rev.\  D {\bf 78} (2008) 015006
  [arXiv:0801.4938 [hep-ph]].

\bibitem{Kl3em}
  V.~Cirigliano, H.~Neufeld and H.~Pichl,
  Eur.\ Phys.\ J.\  C {\bf 35} (2004) 53
  [arXiv:hep-ph/0401173].

  V.~Cirigliano, M.~Knecht, H.~Neufeld, H.~Rupertsberger and P.~Talavera,
  Eur.\ Phys.\ J.\  C {\bf 23} (2002) 121
  [arXiv:hep-ph/0110153].

\bibitem{bsgammaEarly} 
  A.~J.~Buras, M.~Misiak, M.~Munz and S.~Pokorski,
  Nucl.\ Phys.\  B {\bf 424}, 374 (1994)
  [arXiv:hep-ph/9311345].

\bibitem{bsgammaNNLO}
  M.~Misiak and M.~Steinhauser,
  Nucl.\ Phys.\  B {\bf 764}, 62 (2007)
  [arXiv:hep-ph/0609241].

\bibitem{bsgammaHchlimit} 
  M.~Misiak {\it et al.},
  Phys.\ Rev.\ Lett.\  {\bf 98}, 022002 (2007)
  [arXiv:hep-ph/0609232].


\bibitem{SusyBSG}  
  G.~Degrassi, P.~Gambino and P.~Slavich,
  Comput.\ Phys.\ Commun.\  {\bf 179}, 759 (2008)
  [arXiv:0712.3265 [hep-ph]].

\bibitem{bsgammanazila}  
  F.~Mahmoudi,
  arXiv:0808.3144 [hep-ph].

\bibitem{bsgammaQuark} 
 P.~Gambino and M.~Misiak,
  Nucl.\ Phys.\  B {\bf 611}, 338 (2001)
  [arXiv:hep-ph/0104034].

\bibitem{BBbarMixing}
  C.~Q.~Geng and J.~N.~Ng,
  Phys.\ Rev.\  D {\bf 38} (1988) 2857
  [Erratum-ibid.\  D {\bf 41} (1990) 1715].

\bibitem{BBbarMixingNLO}
  J.~Urban, F.~Krauss, U.~Jentschura and G.~Soff,
  Nucl.\ Phys.\  B {\bf 523} (1998) 40
  [arXiv:hep-ph/9710245].

\bibitem{HaberLogan}
  H.~E.~Haber and H.~E.~Logan,
  Phys.\ Rev.\  D {\bf 62}, 015011 (2000)
  [arXiv:hep-ph/9909335].

\bibitem{Field}
 J.~H.~Field,
  Mod.\ Phys.\ Lett.\  A {\bf 13}, 1937 (1998).

\bibitem{Slavich}
 G.~Degrassi and P.~Slavich,
Phys. Rev. D {\bf 81}, 075001 (2010)
[arXiv:1002.1071 [hep-ph]].

\bibitem{ZFITTER}
 A.~B. ~Arbuzov {\it et al.}, 
  Comput. \ Phys.\ Commun.\ {\bf 174} 728-758 (2006). 

\bibitem{elkaffas}
  A.~Wahab El Kaffas, P.~Osland and O.~M.~Ogreid,
  Phys.\ Rev.\  D {\bf 76}, 095001 (2007)
  [arXiv:0706.2997 [hep-ph]].

\bibitem{nirgammaT} 
  Y.~Nir,
  arXiv:0708.1872 [hep-ph].

\bibitem{cleobsg} 
  S.~Chen {\it et al.}  [CLEO Collaboration],
  Phys.\ Rev.\ Lett.\  {\bf 87}, 251807 (2002)
  [arXiv:hep-ex/0108032].

\bibitem{bellebsg}   
  K.~Abe {\it et al.}  [Belle Collaboration],
  AIP Conf.\ Proc.\  {\bf 1078} (2009) 342
  [arXiv:0804.1580 [hep-ex]].

\bibitem{babarbsg}  
B.~Aubert {\it et al.}  [BaBar Collaboration],
Phys. \ Rev. \  D {\bf 72} 052004 (2005). 

\bibitem{buchmullerbsg}
  O.~Buchmuller and H.~Flacher,
  Phys.\ Rev.\  D {\bf 73}, 073008 (2006)
  [arXiv:hep-ph/0507253].

\bibitem{LEPChargedHiggs}
  The LEP Higgs Working Group for Higgs boson searches,
  arXiv:hep-ex/0107031.

\bibitem{lepcdf}
  V.~M.~Abazov {\it et al.}  [D0 Collaboration],
  arXiv:0807.0859 [hep-ex].

  A.~Abulencia {\it et al.}  [CDF Collaboration],
  Phys.\ Rev.\ Lett.\  {\bf 96} (2006) 042003
  [arXiv:hep-ex/0510065].

  D.~Horvath  [OPAL Collaboration],
  Nucl.\ Phys.\  A {\bf 721}, 453 (2003).

  A.~Heister {\it et al.}  [ALEPH Collaboration],
  Phys.\ Lett.\  B {\bf 543} (2002) 1
  [arXiv:hep-ex/0207054].

  P.~Abreu {\it et al.}  [DELPHI Collaboration],
  Phys.\ Lett.\  B {\bf 525}, 17 (2002)
  [arXiv:hep-ex/0201023].

  P.~Achard {\it et al.}  [L3 Collaboration],
  Phys.\ Lett.\  B {\bf 575}, 208 (2003)
  [arXiv:hep-ex/0309056].

\bibitem{scalarFF}
  V.~Bernard, M.~Oertel, E.~Passemar and J.~Stern,
  Phys.\ Lett.\  B {\bf 638} (2006) 480
  [arXiv:hep-ph/0603202].

\bibitem{piK}
  P.~Buettiker, S.~Descotes-Genon and B.~Moussallam,
  Eur.\ Phys.\ J.\  C {\bf 33}, 409 (2004)
  [arXiv:hep-ph/0310283].

\bibitem{taupiK}
  R.~Decker and M.~Finkemeier,
  Phys.\ Lett.\  B {\bf 316}, 403 (1993)
  [arXiv:hep-ph/9307372].

  S.~Banerjee  [BaBar Collaboration],
  arXiv:0811.1429 [hep-ex].

\end{thebibliography}
\end{document}